# Multiattribute Auctions Based on Generalized Additive Independence

**Yagil Engel**                                        YAGILE@IE.TECHNION.AC.IL
*Technion - Israel Institute of Technology*
*Faculty of Industrial Engineering & Management*
*Technion City, Haifa 32000, Israel*

**Michael P. Wellman**                                     WELLMAN@UMICH.EDU
*University of Michigan*
*Division of Computer Science & Engineering*
*2260 Hayward St, Ann Arbor, MI 48109-2121, USA*

## Abstract

We develop multiattribute auctions that accommodate *generalized additive independent* (GAI) preferences. We propose an iterative auction mechanism that maintains prices on potentially overlapping GAI clusters of attributes, thus decreases elicitation and computational burden, and creates an open competition among suppliers over a multidimensional domain. Most significantly, the auction is guaranteed to achieve surplus which approximates optimal welfare up to a small additive factor, under reasonable equilibrium strategies of traders. The main departure of GAI auctions from previous literature is to accommodate *non-additive* trader preferences, hence allowing traders to condition their evaluation of specific attributes on the value of other attributes. At the same time, the GAI structure supports a compact representation of prices, enabling a tractable auction process. We perform a simulation study, demonstrating and quantifying the significant efficiency advantage of more expressive preference modeling. We draw random GAI-structured utility functions with various internal structures, generate additive functions that approximate the GAI utility, and compare the performance of the auctions using the two representations. We find that allowing traders to express existing dependencies among attributes improves the economic efficiency of multiattribute auctions.

## 1. Introduction

Multiattribute trading mechanisms extend traditional, price-only mechanisms by facilitating negotiation over a set of predefined attributes representing various non-price aspects of a deal. Rather than negotiate over a fully specified good or service, a multiattribute mechanism delays commitment to particular configurations until it extracts sufficient information on traders' preferences. For example, a company's procurement department may run a multiattribute auction to select a supplier of hard drives. Supplier offers may be evaluated not only over the price they offer, but also over features such as volume, RPM, access time, latency, transfer rate, and so on. In addition, suppliers may offer contracts differing in terms such as warranty, delivery time, and service.

In order to account for traders' preferences, the auction mechanism must extract evaluative information over a complex domain of multidimensional configurations. Constructing and communicating a complete preference specification can pose a severe burden for even a moderate number of attributes, hence practical multiattribute auctions must either accommodate partial specifications, or support compact expression of preferences assuming some simplified form. By far the most popular multiattribute form to adopt is the simplest: an additive representation where overall value is a linear combination of values associated with each attribute. For example, several recent proposals for iterative multiattribute auctions (Beil & Wein, 2003; Bichler, 2001; David, Azoulay-Schwartz, & Kraus, 2002; Parkes & Kalagnanam, 2005) require additive preference representations.

Such additivity reduces the complexity of preference specification exponentially (compared to the general discrete case), but precludes expression of any interdependencies among the attributes. In practice, however, interdependencies among natural attributes are quite common. For example, the hard-drive buyer may exhibit complementary preferences for volume and access time (since the performance effect is more salient if much data is involved), or may view a strong warranty as a good substitute for high reliability ratings. Similarly, a seller's production characteristics can easily violate additivity, for example if decreasing access time is technically more difficult for higher-capacity drives. In such cases an additive value function may not be able to provide an adequate approximation of real preferences.

On the other hand, fully general models are intractable, and multiattribute preferences typically exhibit *some* structure. Our goal, therefore, is to identify the subtler yet more widely applicable structured representations, and exploit these properties of preferences in trading mechanisms.

We propose an iterative auction mechanism based on just such a flexible preference structure. Our approach is inspired by the design of an iterative multiattribute procurement auction for additive preferences, due to Parkes and Kalagnanam (2005) (PK). PK present two auction designs: the first (NLD) makes no assumptions about traders' preferences, and lets sellers bid on the full multidimensional attribute space. Because NLD maintains an exponential price structure, it is suitable only for small domains. The other auction (AD) assumes additive buyer valuation and seller cost functions. It collects sell bids per attribute level and for a single discount term. The price of a configuration is the sum of the prices of the chosen attribute levels minus the discount.

The auction we propose also supports compact price spaces, albeit for levels of *clusters* of attributes rather than singletons. We employ a preference decomposition based on *generalized additive independence* (GAI), a model flexible enough to accommodate interdependencies to the exact degree of accuracy desired, yet providing a compact functional form to the extent that interdependence can be limited.

First, we build a direct, formally justified link from preference statements over priced outcomes to a generalized additive decomposition of the willingness-to-pay (wtp) function. After laying out this infrastructure, we employ this representation tool for the development of a multiattribute iterative auction mechanism that allows traders to express their complex preferences in GAI format. We then study the auction's allocational, computational, and practical properties. Next, we present a simulation study of our proposed auction mechanism, in order to practically evaluate the economic and computational properties of GAI auctions. We simulate the auctions using random GAI utility functions, including some based on constrained preference structures often exhibited in applications. The simulations let us quantify the benefits of modeling preferences accurately using GAI, in comparison to using an additive approximation. We show that under most circumstances, a GAI auction achieves significantly higher surplus than an auction that uses an additive approximation of preferences.

After providing background on multiattribute preferences and multiattribute auctions (Section 2), we develop new multiattribute structures for wtp functions, supporting generalized additive decompositions (Section 3). We describe our auction mechanism in Section 4, followed by a detailed example in Section 5, and study the mechanism's allocational, computational, and practical properties in Section 6. We present our simulation framework in Section 7, and discuss the experimental results in Section 8.

## 2. Background

In this section we provide essential background on multiattribute preferences (Sections 2.1 and 2.2) and on multiattribute auctions (Section 2.3).

### 2.1 Multiattribute Preferences and Utility

Let $\Theta$ denote the space of possible outcomes, with $\preceq$ a preference relation (weak total order) over $\Theta$. Let $A = \{a_0, \ldots, a_n\}$ be a set of attributes describing $\Theta$. Each attribute $a \in A$ has a domain $D(a)$, so that $\Theta \subseteq \prod_{i=1}^{n} D(a_i)$. Capital letters denote subsets of attributes, small latin letters (with or without numeric subscripts) denote specific attributes, and $\overline{X} = A \setminus X$. $\theta$ (and variations such as $\theta'$ or $\hat{\theta}$) indicate specific outcome in $\Theta$. An instantiation to a subset of attributes $Y$ is denoted using prime signs (as in $Y'$) or numerical superscript (as in $Y^1$). In particular, $Y'$ is a projection on $Y$ of some instantiations $\theta \in \Theta$. To represent an instantiation of subsets $X, Y$ at the same time we use a sequence of instantiation symbols, as in $X^1 Y^2$.

The preference relation $\preceq$ over outcomes is usually represented numerically by a value function $v(\cdot)$ (Keeney & Raiffa, 1976).

**Definition 1 (Value Function).** $v : \Theta \to \Re$ is a *value function* representing $\preceq$ if for any $\theta, \theta' \in \Theta$, $v(\theta) \leq v(\theta')$ iff $\theta \preceq \theta'$.

Clearly, any monotonic transformation of $v(\cdot)$ is also a value function for $\preceq$.

In many cases it is useful to represent, beyond a simple preference order over outcomes, a notion of *strength of preferences*. A value function that expresses strength of preferences is called a *cardinal value function*. A *measurable value function* is a well-established cardinal value framework which posits the existence of a preference order $\preceq^*$ over *pairs of outcomes*. For some $\theta \preceq \theta'$ and $\hat{\theta} \preceq \hat{\theta}'$, the statement $(\theta, \theta') \preceq^* (\hat{\theta}, \hat{\theta}')$ means that the strength of preference of $\hat{\theta}'$ over $\hat{\theta}$ is greater than or equal to that of $\theta'$ over $\theta$. Krantz, Luce, Suppes, and Tversky (1971) establish the set of axioms ensuring the existence of a utility function representing $\preceq^*$.

**Definition 2 (Measurable Value Function).** A *measurable value function* (MVF) is a value function $u : \Theta \to \Re$, such that for any $\theta, \theta', \hat{\theta}, \hat{\theta}' \in \Theta$, for which $\theta \preceq \theta'$ and $\hat{\theta} \preceq \hat{\theta}'$, the following holds:

$$u(\theta') - u(\theta) \leq u(\hat{\theta}') - u(\hat{\theta}) \Leftrightarrow (\theta, \theta') \preceq^* (\hat{\theta}, \hat{\theta}'). \tag{1}$$

Hence the order over *differences in values* of $u(\cdot)$ correspond exactly to the order over preference differences. Note that an MVF can also be used as a value function representing $\preceq$, because $(\theta', \theta) \stackrel{*}{\preceq} (\theta'', \theta)$ iff $\theta' \preceq \theta''$, for any $\theta$.

In auction theory and mechanism design, traders' preferences are usually represented using a quasi-linear value function, such as $v(\theta, p) = u(\theta) + p$, where $p$ represents a monetary outcome.[1] The cardinal value function $u(\theta)$ expresses strength of preference, in that the difference $u(\theta') - u(\theta'')$ corresponds to the additional amount a trader is willing to pay for $\theta'$ relative to $\theta''$. For example, if $\theta'$ represents a red Mercedes with a sunroof, and $\theta''$ denotes a blue Subaru with no sunroof, then $u(\theta') - u(\theta'')$ is the strength of preference for the Mercedes configuration over the Subaru. If the Mercedes costs $p'$ and the Subaru $p''$, then according to $v(\theta, p)$ the trader prefers the Mercedes deal iff $u(\theta') - u(\theta'') \geq p' - p''$.

In fact, $u(\cdot)$ can be easily shown to be an MVF, where the preference differences correspond to differences in willingness-to-pay (Engel & Wellman, 2007). For this reason, we use MVF as the basis for utility in this work, and assume that traders' willingness-to-pay (wtp) functions constitute an MVF.

Reasoning over full outcomes is hard in several ways. Most notably, it is difficult for humans to compare outcomes over many dimensions, and complex for machines to store and analyze preferences over a number of outcomes that is exponential in the number of attributes. It is therefore useful to consider preferences over the joint product of some $Y \subset A$, considering the rest of the attributes $\overline{Y}$ fixed on some predefined values. Such an order is also often referred to as a *ceteris paribus* preference order—one partial outcome is preferred to another *all else being equal*.

**Definition 3 (Conditional Preference).** Partial outcome $Y^2$ is *conditionally preferred* to partial outcome $Y^1$ given $\overline{Y}'$, if $Y^1 \overline{Y}' \preceq Y^2 \overline{Y}'$. The conditional preference order over $Y$ given $\overline{Y}'$ is denoted by $\preceq_{\overline{Y}'}$, hence $Y^1 \overline{Y}' \preceq Y^2 \overline{Y}'$ is abbreviated $Y^1 \preceq_{\overline{Y}'} Y^2$.

In general, conditional preferences may depend on the particular assignment chosen for the rest of the attributes. More precisely, if $Y^1 \prec_{\overline{Y}'} Y^2$, we could still find that $Y^2 \prec_{\overline{Y}''} Y^1$ for some $\overline{Y}'' \neq \overline{Y}'$. When this is the case, one needs to maintain both conditional preference orders $\preceq_{\overline{Y}'}$ and $\preceq_{\overline{Y}''}$, and hence in general this scheme might not yield any computational benefits. Fortunately, in many cases one can identify subsets $Y$ for which this preference reversal does not occur, that is the preference order over $Y$ is invariant to the instantiation of $\overline{Y}$.

**Definition 4 (Preferential Independence).** $Y$ is *preferential independent (PI)* of $\overline{Y}$, written PI($Y, \overline{Y}$), if for any $Y^1$ and $Y^2$, and for any $\overline{Y}', \overline{Y}''$, we have $Y^1 \preceq_{\overline{Y}'} Y^2$ iff $Y^1 \preceq_{\overline{Y}''} Y^2$.

*First-order preferential independence* (FOPI), independence of a single attribute from the rest, is a natural assumption in many domains. For example, in typical purchase decisions greater quantity or higher quality is more desirable regardless of the assignments to other attributes. Preferential independence of higher order, however, requires invariance of the *tradeoffs among some attributes* with respect to variation in others, a more stringent independence condition. The MPI condition, defined below, is over the global set of attributes $A$, and requires *all* possible subsets to be PI.

---

1. We use the term *trader* when referring to either buyers or sellers.

**Definition 5** (**Mutual Preferential Independence**). Attributes $A$ are *mutually preferential independent* (MPI) iff for all $Y \subset A$, $PI(Y, \overline{Y})$.

Preferential independence can greatly simplify the form of $v$.

**Theorem 1** (Debreu, 1959). *A preference order over set of attributes $A$ can be represented by an additive value function*

$$v(a_1, \ldots, a_n) = \sum_{i=1}^{n} v_i(a_i),$$

*iff $A$ is mutually preferential independent.*

Dyer and Sarin (1979) extend additivity theory to MVF, and specify the conditions under which $u(\cdot)$ as well has an additive structure as above. Effectively, additive forms used in trading mechanisms assume MPI over the full set of attributes, including the money attribute. Intuitively that means willingness-to-pay for levels of an attribute or attributes cannot be affected by the instantiation of other attributes. This sweeping condition rarely holds in practice (Von Winterfeldt & Edwards, 1986). Therefore, recent AI literature often relaxes the MPI assumption by imposing additivity only with respect to *subsets* of attributes which may *overlap*.

**Definition 6** (**Generalized Additive Independence**). Let $I_1, \ldots, I_g$ be (not necessarily disjoint) subsets of $A$, such that $\bigcup_{i=1}^{g} I_i = A$. The *elements* $I_1, \ldots, I_g$ are called *generalized additive independent* (GAI) if there exist functions $f_1, \ldots, f_g$ such that,

$$u(a_1, \ldots, a_n) = \sum_{r=1}^{g} f_r(I_r). \tag{2}$$

## 2.2 Related Work on Generalized Independence

Our definition of GAI is somewhat nonstandard. Most literature defines a GAI condition for the *expected utility* function (von Neumann & Morgenstern, 1944). In that well-known model, a particular choice results in a *lottery*, that is a probability distribution over outcomes. The expected utility function represents a complete preference order over lotteries. Informally, the GAI definition requires preferences on lotteries over $\Theta$ to depend only on the margins over the subsets $I_1, \ldots, I_g$. The form of Eq. (2) is a result of that definition, obtained by Fishburn (1967). Fishburn not only introduces the functional decomposition, but also provides a well-defined form for the functional constituents $f_1, \ldots, f_g$. Graphical models and elicitation procedures for GAI decomposable utility were developed within the expected utility framework (Bacchus & Grove, 1995; Boutilier, Bacchus, & Brafman, 2001; Gonzales & Perny, 2004; Braziunas & Boutilier, 2005). In addition, generalized additive utility models have been employed by Hyafil and Boutilier (2006) as an aid in direct revelation mechanisms, and by Robu, Somefun, and La Poutré (2005) for opponent modeling in bilateral multi-item negotiation.

Bacchus and Grove (1995), who in fact coined the term GAI, show that the decomposition can also be obtained as a result of a collection of local, easier to detect, *binary* independence conditions. More specifically, they rely on a form called *conditional additive independence*, which, informally, corresponds to a GAI decomposition limited to two (overlapping) subsets $X \subset A$ and $Y \subset A$. They prove that this condition can be expressed as a separation criterion on a graph whose nodes correspond to $A$, by the means of a *perfect map* (Pearl, 1988). Crucially, the utility function decomposes to GAI form over lower dimensional functions, each defined on a maximal clique of the graph. When combined with Fishburn's work, their result provides a well-defined functional form that can be obtained from a collection of conditional additive independence conditions. This result relies on the form of lotteries as a basis for the utility function and the independence conditions.

The expression of willingness-to-pay requires a cardinal measure of preferences, yet without uncertainty, there is no need for an expected utility representation. We therefore invoke the MVF framework, and in Section 3, build on the additive decompositions for MVF developed by Dyer and Sarin (1979) to develop multiattribute preference structures for wtp. This development enables us to follow the footsteps of Fishburn (1967) and Bacchus and Grove (1995) and show that a well-defined GAI form for MVF can also be obtained using a collection of easy-to-detect binary independence conditions.

### 2.3 Multiattribute Auctions

The distinguishing feature of a multiattribute auction is that the goods are defined by vectors of *attributes*. As above, we use $A$ to denote a set of attributes describing the domain $\Theta$. A *configuration* is a particular attribute vector, $\theta \in \Theta$. Multiattribute auctions are used primarily for procurement, as part of strategic sourcing processes (Sandholm, 2007). In the procurement model there is a single buyer, who has a utility function (representing willingness-to-pay) $u_b(\theta)$ for purchasing $\theta \in \Theta$. There are $m$ sellers $s_1, \ldots, s_m$ with utility functions $c_i : \Theta \to \Re$, representing the cost for $s_i$ to supply configurations in $\Theta$ to the buyer.

**Definition 7 (Multiattribute Allocation Problem).** The *multiattribute allocation problem* (Parkes & Kalagnanam, 2005) is:

$$MAP = \max_{i \in \{1, \ldots, m\}, \theta \in \Theta} u_b(\theta) - c_i(\theta). \tag{3}$$

An allocation $(s_i^*, \theta^*)$ solving $MAP$ is said to maximize the *surplus* of the procurement problem.

$MAP$ can be decomposed to two subproblems: first find the most efficient configuration *for each trader*, and then find the trader whose efficient configuration yields the highest surplus. We call the first part the *multiattribute matching problem* (Engel, Wellman, & Lochner, 2006).

**Definition 8 (Multiattribute Matching Problem).** The *multiattribute matching problem* (MMP) for a buyer $b$ and a seller $s_i$ is:

$$MMP(b, s_i) = \arg\max_{\theta \in \Theta} u_b(\theta) - c_i(\theta).$$

We also call a configuration $\theta^*$ selected by $MMP(b, s_i)$ a *bilaterally efficient* configuration for $s_i$.

Most of the theoretical work on surplus-maximizing multiattribute auctions relates in some way to the foundational work by Che (1993). In Che's model, the good or service is characterized by a single quality attribute, and each seller has an independent private cost function over quality. The buyer announces a scoring rule to the sellers, by which price-quality offers will be evaluated. Che suggests several types of auctions, including the *second-score auction*, where the seller bidding highest score wins, and must provide a combination of price and quality that achieves the second-best score. In the second-score mechanism, bidding truthfully is an equilibrium in dominant strategies. In particular, Che shows that sellers bid on the quality that maximizes the difference between the buyer's scoring rule and their own cost function; in other words, on the respective MMP solution. Branco (1997) generalizes Che's model and some of his results to correlated costs.

This basic model was later generalized by several authors to account explicitly for multiple quality attributes, and usually restricting the scoring rule to be additive over the attributes (Bichler, 2001; David et al., 2002). Vulkan and Jennings (2000) suggest a modified version of English auctions (iterative auctions that require new bids to increment over current bid price) under which bidders are required to improve current score, rather than price. Sandholm and Suri (2006) consider the incorporation of non-price attributes in multi-item (combinatorial) auctions.

The literature surveyed above emphasizes that auctions require the buyer to reveal a scoring function prior to bidding. In order to achieve economic efficiency, this scoring function must convey the buyer's full utility function $u_b(\cdot)$. This is a major obstacle to practical adaption of these mechanisms. Procurement auctions are rarely an isolated event, and the buyer-supplier relationships usually evolve and change over time, during which suppliers may retain some market power, and take advantage of the information revealed by the buyer. Events are sometimes conducted on a recurrent basis, and several events may be conducted for related goods with correlated valuations. In addition, the buyer may wish to keep secret the way her utility may be discriminating for or against particular suppliers (Koppius, 2002).

As noted in Section 1, Parkes and Kalagnanam (2005) suggest an alternative approach, which employs prices over the space of configurations to drive traders to the efficient configurations. Auction NLD maintains a price for each $\theta \in \Theta$, and sellers bid for such full configuration in each round. Auction AD maintains a price for each level $a_i' \in D(a_i)$. Prices are initially set to be high. In each round, sellers bid on a particular value for each attribute, and the auction selects a set of levels (again, per attribute) which are myopically *buyer preferred* in that round, that is, approximately maximize the buyer's utility with respect to current prices. In addition, the auction maintains a discount factor that is applied to ensure that a single seller is eventually selected. The price of a configuration is defined as the sum of the prices of the chosen attribute levels minus the discount. After each round, prices of particular levels of particular attributes are decremented by a constant $\epsilon$, according to a set of price change rules, ensuring that the auction ultimately converges to an efficient solution.

Both auctions are shown to obtain optimal surplus (up to $\epsilon$-proportional error), when all sellers bid myopically rather than strategically (we define this concept formally in Section 6.1). The myopic behavior is shown to be an ex-post Nash equilibrium. Auction NLD is fully expressive but not tractable when the number of attributes is large. Auction AD is computationally efficient, but its expressiveness is limited to additive preferences (see discussion following Theorem 1). When traders' preferences are not additive, the welfare achieved by the auction is not necessarily optimal; that is, it does solve MAP optimally, but with respect to inaccurate utility functions. Moreover, it is not clear how this lack of expressiveness may affect the incentives of traders to act strategically.

Theoretically, one could also use the well-known Vickrey-Clake-Grove (VCG) mechanism. Parkes and Kalagnanam define the *sell-side* VCG mechanism: all traders submit their full utility or cost functions, MAP is solved by the auction engine, and the winning seller pays according to her VCG price (definition of this pricing is provided in Section 6.1). In such an auction, traders can be allowed to use any compact preference structure, including GAI. However, this scheme suffers from the same disadvantages as any of the proposals that require full revelation of utility.

To summarize, no previously suggested surplus-maximizing multiattribute procurement auction is at the same time *expressive* (accommodates interdependencies between attributes), *tractable* (its computations do not depend on the fully exponential domain), and *preserving of the buyer's private information*, meaning (minimally) that it does not require the buyer to reveal a full utility function before extracting any bids from sellers. Our proposed mechanism, as we show theoretically and using simulations, possesses attractive properties on all these criteria.

## 3. Detection of GAI Structure for Measurable Value Functions

In this section we provide the basis for the application of GAI decomposition in procurement problems. In Section 3.1 we show how GAI can be obtained as a collection of local, weaker conditions which are based on invariance of willingness-to-pay. In Section 3.2 we use an example to demonstrate how this process can be used in procurement problems.

### 3.1 Difference Independence and GAI

Dyer and Sarin (1979) introduce for measurable value an analog to additive independence, called *difference independence*. Our first step is to introduce a conditional generalization of their definition.

**Definition 9 (Conditional Difference Independence).** Let $X, Y \subset A$ and $X \cap Y = \emptyset$, and define $Z = A \setminus X \cup Y$. $X$ is *conditionally difference independent* of $Y$, denoted as $\text{CDI}(X, Y)$, if for any $Z' \in D(Z)$, and for any $X^1, X^2 \in D(X), Y^1, Y^2 \in D(Y)$,

$$(X^1 Y^1 Z', X^2 Y^1 Z') \sim (X^1 Y^2 Z', X^2 Y^2 Z'), \tag{4}$$

where the symbol $\sim$ indicates that $\overset{\cdot}{\preceq}$ and $\overset{\cdot}{\succeq}$ both hold.

By the definition of MVFs (1), the CDI condition (4) can be expressed equivalently in terms of measurable value:

$$u(X^1 Y^1 Z') - u(X^2 Y^1 Z') = u(X^1 Y^2 Z') - u(X^2 Y^2 Z')$$

This condition states that the value, or willingness-to-pay, for a change in the assignment to $X$ does not depend on the current assignment of $Y$, for any fixed value of $Z$.

A CDI condition leads to a convenient decomposition of the MVF.

**Lemma 2.** *Let $u(A)$ be an MVF representing preference differences, with $X, Y, Z$ as specified in Definition 9. Then* $\mathrm{CDI}(X, Y)$ *iff*

$$u(A) = u(X^0, Y, Z) + u(X, Y^0, Z) - u(X^0, Y^0, Z),$$

*for any arbitrary instantiations $X^0, Y^0$.*

With a single CDI condition, we can therefore replace the $n$-ary function $u(X, Y, Z)$ with two lower-dimensional functions $u(X^0, Y, Z)$ and $u(X, Y^0, Z)$. It is reasonable to assume that one can apply more CDI conditions to further decompose the resulting functions. In order to take full advantage of *all* existing CDI conditions, we introduce the notion of a dependency graph, which is a simplification of the concept of *perfect map* mentioned in Section 2.2.

**Definition 10** (**Dependency Graph**). Let $\mathcal{S}$ denote a set, and $\mathcal{R}$ a binary relation over $2^{\mathcal{S}}$. Then a graph $G = (\mathcal{S}, E)$ is a dependency graph for $\mathcal{R}$ if for any $S_1, S_2 \subset \mathcal{S}$, it holds that $(S_1, S_2) \in \mathcal{R}$ iff for any $a_1 \in S_1$ and $a_2 \in S_2$, $(a_1, a_2) \notin E$.

Hence the dependency graph expresses $\mathcal{R}$ as a separation criterion; two subsets have a direct connection iff they are *not* in $\mathcal{R}$. A dependency graph for CDI can be constructed simply by removing any edge $(a_1, a_2)$ for which $\mathrm{CDI}(\{a_1\}, \{a_2\})$; this is because $\mathrm{CDI}(S_1, S_2)$ holds iff $\mathrm{CDI}(\{a_1\}, \{a_2\})$ holds for any $a_1 \in S_1$ and $a_2 \in S_2$. We use the term *CDI map* for a dependency graph induced by a CDI relation.

The next theorem links the CDI condition, the CDI map, and a GAI decomposition over $A$. In fact, it establishes that the functional constituents of GAI decomposition for MVF are the same as the functional constituents of GAI decomposition for the expected utility model, as defined by Fishburn (1967). We adopt the following conventional notation. Let $(a_1^0, \ldots, a_n^0)$ be a predefined vector called the *reference outcome*. For any $I \subseteq A$, the function $u([I])$ stands for the projection of $u(A)$ to $I$ where the rest of the attributes are fixed at their reference levels.

**Theorem 3** (**CDI-GAI Theorem**). *Let $G = (A, E)$ be a CDI map for $A$, and $\{I_1, \ldots, I_g\}$ a set of (overlapping) maximal cliques of $G$. Then*

$$u(A) = \sum_{r=1}^{g} f_r(I_r), \tag{5}$$

*where*

$$
\begin{aligned}
f_1 &= u([I_1]), \text{ and} \\
\text{for } r = 2, \ldots, g, \quad f_r &= u([I_r]) + \sum_{j=1}^{r-1} (-1)^j \sum_{1 \le i_1 < \cdots < i_j < r} u([\bigcap_{s=1}^{j} I_{i_s} \cap I_r]).
\end{aligned}
\tag{6}
$$

As a small example, Table 1 exhibits a utility function $u(x_1, x_2, x_3)$. Each of the three attributes has a boolean domain, that is $D(x_i) = \{0, 1\}$. Let $x_i^0$ and $x_i^1$ denote the assignments 0 and 1 (respectively) to $x_i$. We first observe that $\mathrm{CDI}(\{x_1\}, \{x_3\})$ holds because:[2]

---

2. Note that $x_2^0$ and $x_2^1$ correspond to $Z'$ in Definition 9.

| $x_1$ | $x_2$ | $x_3$ | $u(x_1, x_2, x_3)$ | $u(x_1, x_2, x_3^0)$ | $u(x_1^0, x_2, x_3)$ | $u(x_1^0, x_2, x_3^0)$ | $u_1(I_1)$ | $u_2(I_2)$ |
|---|---|---|---|---|---|---|---|---|
| 0 | 0 | 0 | 0 | 0 | 0 | 0 | 0 | 0 |
| 1 | 0 | 0 | 5 | 5 | 0 | 0 | 5 | 0 |
| 0 | 1 | 0 | 2 | 2 | 2 | 2 | 2 | 0 |
| 1 | 1 | 0 | 6 | 6 | 2 | 2 | 6 | 0 |
| 0 | 0 | 1 | 3 | 0 | 3 | 0 | 0 | 3 |
| 1 | 0 | 1 | 8 | 5 | 3 | 0 | 5 | 3 |
| 0 | 1 | 1 | 7 | 2 | 7 | 2 | 2 | 5 |
| 1 | 1 | 1 | 11 | 6 | 7 | 2 | 6 | 5 |

Table 1: A utility function over three attributes, decomposable via GAI into the sum of two functions of two attributes each. $u_1(\cdot)$ depends only on $\{x_1, x_2\}$ and $u_2(\cdot)$ depends only on $\{x_2, x_3\}$.

1. The utility difference on values of $x_1$ given $x_2^0$ is 5, for both $x_3^0$ and $x_3^1$. More explicitly, $u(x_1^1, x_2^0, x_3^0) - u(x_1^0, x_2^0, x_3^0) = 5 - 0 = 5$, and $u(x_1^1, x_2^0, x_3^1) - u(x_1^0, x_2^0, x_3^1) = 8 - 3 = 5$.

2. Similarly, the difference on $x_1$ given $x_2^1$ is 4, for both $x_3^0$ and $x_3^1$.

Though $x_1$ and $x_3$ are CDI of each other, we see that both depend on $x_2$. For example, the differences mentioned above for $x_1$ are 5 and 4 given $x_2^0$ and $x_2^1$ (respectively), hence the difference on $x_1$ given fixed $x_3$ depends on the value of $x_2$. The CDI map for this example therefore includes an edge $(x_1, x_2)$ and an edge $(x_2, x_3)$. The maximal cliques of this graph are $I_1 = \{x_1, x_2\}$ and $I_2 = \{x_2, x_3\}$.

To obtain the numeric decomposition, we first define $(x_1^0, x_2^0, x_3^0)$ as reference values. Next, from (6), we get $u_1(I_1) = u([I_1]) = u(x_1, x_2, x_3^0)$ and $u_2(I_2) = u([I_2]) - u([I_1 \cap I_2]) = u(x_1^0, x_2, x_3) - u(x_1^0, x_2, x_3^0)$. The functions involved are given in Table 1. Note that the fifth and sixth columns are obtained from the appropriate values of the fourth column; for example, $u(x_1^0, x_2, x_3)$ for the line $x_1 = 1, x_2 = 1, x_3 = 0$ is the value $u(x_1, x_2, x_3)$ in the line $x_1 = 0, x_2 = 1, x_3 = 0$. It is easy to verify that indeed $u(x, y, z) = u_1(I_1) + u_2(I_2)$.

The CDI-GAI Theorem provides an operational form of GAI, by establishing a GAI decomposition that can be obtained from a collection of simple CDI conditions. The assumption or detection of CDI conditions can be performed incrementally, until the MVF is decomposed to a reasonable dimension. The CDI conditions, in turn, based as they are on invariance of preference differences, are relatively intuitive to detect. This is particularly true when the differences carry a direct interpretation, as in the case of willingness-to-pay: we can check invariance of the monetary amount a buyer is willing to pay to get one outcome over the other.

The GAI decomposition can be depicted graphically using a clique graph of the CDI map, that is, a graph whose nodes correspond to maximal cliques of the CDI map. For our purposes it is convenient to use a particular clique graph called a *tree decomposition* (or *junction tree*). We introduce this well-known concept, and discuss its implications for GAI representation.

**Definition 11 (Tree Decomposition).** A *tree decomposition* for a graph $G = (N, E)$ is a pair $(T, \mathcal{I})$, where $T = (\Psi, \mathcal{E})$ is an acyclic graph, $\mathcal{I} = \{I_i \mid i \in \Psi\}$ is a collection of subsets of $N$, each corresponding to a node in $T$, and (i) $\bigcup_{i \in \mathcal{I}} I_i = N$, (ii) for each edge $(n_1, n_2) \in E$, there exists $I_i$ such that $n_1, n_2 \in I_i$, and (iii) (running intersection) for any $i, j, k \in \Psi$, if $j$ is on the path from $i$ to $k$ in $T$ then $I_i \cap I_k \subseteq I_j$.

| term | Meaning | Reference |
|------|---------|-----------|
| MAP | Multiattribute Allocation Problem | (Parkes & Kalagnanam, 2005) |
| MMP | Multiattribute Matching Problem | (Engel et al., 2006) |
| MVF | Measurable Value Function | (Dyer & Sarin, 1979) |
| PI | Preferential Independence | (Keeney & Raiffa, 1976) |
| FOPI | First-Order Preferential Independence | |
| MPI | Mutual Preferential Independence | (Keeney & Raiffa, 1976) |
| GAI | Generalized Additive Independence | (Bacchus & Grove, 1995) |
| CDI | Conditional Difference Independence | |
| CDI map | graph whose separation criterion is CDI | cf. (Bacchus & Grove, 1995) |
| GAI tree | tree decomposition of a CDI map | cf. (Gonzales & Perny, 2004) |

Table 2: Acronym terms, with references to related literature. Empty references indicate terms introduced in this work. The terms are arranged according to topics: (i) multiattribute economic problems, (ii) independence relations, (iii) graphical concepts.

Any graph can be tree-decomposed, typically in more than one way. For example, there can be a single node in $\mathcal{I}$. The *width* of a tree decomposition is $\max_{i \in \mathcal{I}} |I_i| - 1$, and the *treewidth* of a graph is the minimum width among all its possible tree decompositions.

It is easy to show that any maximal clique of $G$ is contained within some $i \in \mathcal{I}$. Therefore, by Theorem 3, a utility function decomposes additively over the subsets $\mathcal{I} = \{I_i \mid i \in \Psi\}$, where $T = (\Psi, \mathcal{E})$ is a tree decomposition of the CDI map. The notion of *GAI tree* is adapted from the work of Gonzales and Perny (2004), who introduce GAI graphical models for the expected utility framework.

**Definition 12 (GAI Tree).** A GAI tree for $u(\cdot)$ is a tree decomposition of the CDI map of $u(\cdot)$.

We therefore refer to the elements $I_1, \ldots, I_g$ of a GAI decomposition as the set $\mathcal{I}$ of a tree decomposition. The next subsection provides a qualitative example of the CDI concept, its dependency graph, and corresponding GAI tree.

The results of this section lay out the foundations for using GAI decomposition in multiattribute trading mechanisms. The results generalize additive MVF theory, and justify the application of methods developed under the expected utility framework (Bacchus & Grove, 1995; Boutilier et al., 2001; Gonzales & Perny, 2004; Braziunas & Boutilier, 2005) to representation of monetary value under certainty. Table 2 summarizes the acronym terminology introduced up to this point.

## 3.2 Employing GAI in Procurement

In this section we demonstrate the process of obtaining a GAI decomposition from the collection of CDI conditions. In addition, this example is used to motivate our approach in comparison to the work of Parkes and Kalagnanam (2005). Consider a procurement department that wishes to purchase new hard drives (HD) for the desktops of a large number of employees. The buyer cares about several characteristics (attributes) of the hard drives and the particular terms of the procurement contract. Each attribute is listed with a designated attribute name (the first letter), and its domain. In some cases (e.g., attribute $I$) we use arbitrary symbols to represent domain elements, abstracting from the meaningful interpretation they are assumed to have in context.

**RPM (R)** 3600, 4200, 5400 RPM

**Transfer rate (T)** 3.4, 4.3, 5.7 MBS

**Volume (V)** 60, 80, 120, 160 GB

**Supplier ranking (S)** 1, 2, 3, 4, 5

**Quality rating (Q)** (of the HD brand) 1, 2, 3, 4, 5

**Delivery time (D)** 10, 15, 20, 25, 30, 35 days

**Warranty (W)** 1, 2, 3 years

**Insurance (I)** (for the case the deal is signed but not implemented) $\alpha_1$, $\alpha_2$, $\alpha_3$

**Payment timeline (P)** 10, 30, 90 days

**Compatibility (C)** (with existing hardware and software) $\beta_1$, $\beta_2$, $\beta_3$

Consider, for example, the pair of attributes Quality and Warranty. The value of warranty is different for different values of quality; it is higher when the quality is known to be low, and lower when the quality is known to be high. The two attributes therefore depend on each other. Furthermore, we might expect that Volume complements both Quality and Warranty. Larger hard drives are more prone to failures, making the quality and warranty more valuable. Similarly, there is interdependence between Supplier ranking and the contract insurance we buy, and between Supplier ranking and the warranty the supplier provides. Other reasonable dependencies are among Delivery, Insurance, and Payment timeline (e.g., later delivery requires better insurance, later payment reduces the need for insurance), and between Volume to the RPM and Transfer rate. Preferences over compatibility may not depend on any other attribute. The corresponding CDI map is depicted in Figure 1a. As described in Section 2.1, the utility function decomposes over the elements of a tree decomposition of the CDI map. Such a tree decomposition is depicted in Figure 1b. In this example the set of elements of the tree decomposition correspond exactly to the maximal cliques of the CDI map. In general the tree decomposition might include supersets of the maximal cliques, but the decomposition can obviously be maintained over the supersets as well.

Non-additive traders, if required to deal with an additive price space as in auction AD (Parkes & Kalagnanam, 2005), face an *exposure problem*, somewhat analogous to traders with combinatorial preferences that participate in simultaneous auctions (Wellman, Osepayshvili, MacKie-Mason, & Reeves, 2008). Essentially, the problem can manifest itself in two ways. One type of exposure occurs from one auction round to another, as in the following two-attribute example. A seller's conditional preference order over an attribute $a$ may be optimized at an assignment $a^1$ given that the other attribute $b$ is at $b^1$, but if the assignment of $b$ changes, $a^1$ may become arbitrarily suboptimal. Therefore bidding $a^1$ and $b^1$ may result in a poor allocation if the seller is outbid on $b^1$ (and thus must resort to another assignment) but left winning $a^1$. The second exposure occurs in any single round of the auction, if a trader bids on multiple configurations. For example, suppose configurations $(a^1, b^1)$ and $(a^2, b^2)$ are both optimal at the current prices. Because bids are collected independently for each attribute, a trader bidding on both may end up with configuration $(a^1, b^2)$, which again, may be arbitrarily suboptimal.

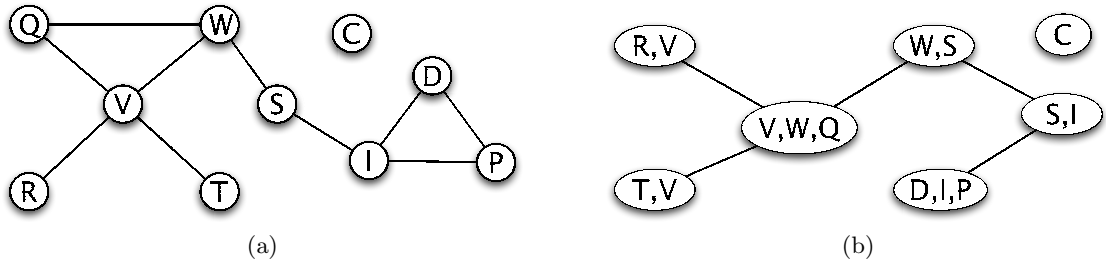

Figure 1: HD procurement problem: (a) CDI map, (b) GAI tree.

We can prevent exposure on the sellers' part by taking simple measures in the auction design. First, bids are collected anew each round, independently of previous rounds, hence the first problem is avoided. Sellers can likewise avoid the second problem by limiting themselves to bid on one configuration per round.

On the buyer's side, this solution does not work because we require the buyer to bid a full set of optimal configurations in each round, in order to ensure the auction's convergence (this becomes clearer in Section 6.1). To prevent buyer exposure, our auction design structures prices according to the buyer's preferences, and the traders bid on clusters of interdependent attributes. In terms of the example above, if $a$ and $b$ are interdependent (meaning CDI($\{a\}, \{b\}$) does not hold), we should be able to bid on the cluster $ab$. If $b$ in turn depends on $c$, we need another cluster $bc$. This is still simpler than a general pricing structure that solicits bids for the cluster $abc$. More generally, we find all reasonable CDI conditions which are correct for the buyer, obtain the corresponding GAI tree decomposition, and solicit bids for clusters of attributes corresponding to these GAI elements. In Section 4, we describe our auction design in detail, along with an example in Section 5. In Section 6.1, we prove that the auction terminates with an (approximately) optimal solution to MAP.

## 4. GAI Auctions

Before introducing our auction design, we reiterate our model and notation, and provide a definition that facilitates the auction presentation.

### 4.1 Notations and Definitions

In the procurement setting, a single buyer wishes to procure a single good, in some configuration $\theta \in \Theta$ from one of the candidate sellers $s_1, \ldots, s_m$. The buyer has some private valuation function $u_b : \Theta \to \Re^+$, and similarly each seller $s_i$ has a private cost function, $c_i$. Both $u_b(\cdot)$ and $c_i(\cdot)$ are MVFs, for which utility differences express differences in willingness-to-pay, as explained in Section 2.1. Assume that the buyer's preferences are reflected in a GAI structure $I_1, \ldots, I_g$. We call an assignment to GAI element $I_r$ a *sub-configuration*. We use $\theta_r$ to denote the sub-configuration formed by projecting configuration $\theta$ to element $I_r$.

**Definition 13 (Consistent Cover).** A collection of sub-configurations $\{\alpha_1, \ldots, \alpha_g\}$, where for each $r \in \{1, \ldots, g\}$, $\alpha_r$ is an instantiation of $I_r$, is a *consistent cover* if for any $r, r' \in \{1, \ldots, g\}$, and any attribute $a_j \in I_r \cap I_{r'}$, $\alpha_r$ and $\alpha_{r'}$ agree on the assignment to $a_j$.

In words, a consistent cover is a collection of sub-configurations from which we can compose a valid configuration. A collection $\{\alpha_1, \ldots, \alpha_g\}$ which is a consistent cover can equivalently be considered a configuration, which we denote by $(\alpha_1, \ldots, \alpha_g)$. For example, consider a good with three attributes: $a, b, c$. Each attribute's domain has two possible assignments (e.g., $\{a^1, a^2\}$ is the domain of $a$). Let the GAI structure be $I_1 = \{a, b\}, I_2 = \{b, c\}$. Here, sub-configurations are assignments of the form $a^1 b^1$, $a^1 b^2$, $b^1 c^1$, and so on. The set of sub-configurations $\{a^1 b^1, b^1 c^1\}$ is a consistent cover, corresponding to the configuration $a^1 b^1 c^1$. In contrast, the set $\{a^1 b^1, b^2 c^1\}$ is inconsistent.

### 4.2 The GAI Auction

We define an iterative, descending-price multiattribute auction that maintains a GAI pricing structure: that is, in any round $t$, there is a price $p^t(\cdot)$, corresponding to each sub-configuration of each GAI element. The price $p^t(\theta)$ of a configuration $\theta$ at round $t$ is defined in terms of the sub-configuration prices and a global discount term $\Delta$,

$$p^t(\theta) = \sum_{r=1}^{g} p^t(\theta_r) - \Delta. \tag{7}$$

Importantly, the elements $\theta_r$ may refer to overlapping attributes. Bidders submit *sub-bids* on sub-configurations and on the global discount $\Delta$.[3] Sub-bids are submitted in each round and they expire in the next round. A sub-bid in round $t$ for configuration $\theta_r$ is automatically assigned the price $p^t(\theta_r)$. The set of *full bids* of a seller contains all consistent covers that can be generated from that seller's current set of sub-bids. The existence of a full bid over a configuration $\theta$ represents the seller's willingness to accept the price $p^t(\theta)$ for supplying $\theta$.

At the start of the auction, the buyer reports (to the auction, not to sellers) a complete valuation function $u_b(\cdot)$. Under GAI, this can be expressed in decomposed form (6) with local functions $(f_{b,1}, \ldots, f_{b,g})$, such that $u_b(\theta) = \sum_r f_{b,r}(\theta_r)$. The initial prices of sub-configurations are set at some level above the buyer's valuations, that is, $p^1(\theta_r) > f_{b,r}(\theta_r)$ for all $\theta_r$. The discount $\Delta$ is initialized to zero. The auction has the dynamics of a *descending*

---

3. The discount term could be replaced with a uniform price reduction across all sub-configurations.

*clock auction*: at each round $t$, bids are collected for current prices and then prices are reduced according to price rules. A seller is considered *active* in a round if the set of sub-bids she submitted contains at least one full bid. In round $t > 1$, only sellers who were active in round $t - 1$ are allowed to participate, and the auction terminates when no more than a single seller is active. We denote the set of sub-bids submitted by $s_i$ by $\mathcal{B}_i^t$, and the corresponding set of full bids is

$$B_i^t = \{\theta = (\theta_1, \ldots, \theta_g) \in \Theta \mid \{\theta_1, \ldots, \theta_g\} \subseteq \mathcal{B}_i^t\}.$$

In the example of Section 4.1, a seller could submit sub-bids on a set of sub-configurations such as $\{a^1 b^1, b^1 c^1\}$, and that combines to a full bid on $a^1 b^1 c^1$.

The auction proceeds in two phases. In the first phase (A), at each round $t$ the auction computes a set of *buyer-preferred* sub-configurations $\mathcal{M}^t$: those sub-configurations that are part of a configuration which is within $\epsilon$ of being profit-maximizing for the buyer at the current prices. Formally, we first define the buyer profit from a configuration $\theta$ as[4]

$$\pi_b^t(\theta) = u_b(\theta) - p^t(\theta).$$

The buyer-preferred set of sub-configurations is then defined by:

$$\mathcal{M}^t = \{\theta_r \mid \pi_b^t(\theta) \geq \max_{\theta' \in \Theta} \pi_b^t(\theta') - \epsilon, r = 1, \ldots, g\}.$$

In Section 6.2 we show how $\mathcal{M}^t$ can be computed efficiently. We stress that though $\mathcal{M}^t$ is a set of sub-configurations, the criterion for selecting them is based on the profit over full configurations. Profits over individual sub-configurations are meaningless outside the context of configurations.

In Phase A, the auction adjusts prices after each round, reducing the price of every sub-configuration that has received a bid but is not in the buyer's preferred set. Let $\epsilon$ be the prespecified price decrement parameter. Specifically, the Phase A price change rule is applied to all $\theta_r \in \bigcup_{i=1}^m B_i^t \setminus \mathcal{M}^t$:

$$p^{t+1}(\theta_r) \leftarrow p^t(\theta_r) - \frac{\epsilon}{g}. \qquad [\text{A}]$$

Let $M^t$ denote the set of configurations that are consistent covers in $\mathcal{M}^t$:

$$M^t = \{\theta = (\theta_1, \ldots, \theta_g) \in \Theta \mid \{\theta_1, \ldots, \theta_g\} \subseteq \mathcal{M}^t\}.$$

The auction switches to Phase B when all active sellers have at least one full bid in the buyer's preferred set:

$$\forall i.\ B_i^t = \emptyset \lor B_i^t \cap M^t \neq \emptyset. \qquad [\text{SWITCH}]$$

Let $T$ be the round at which [SWITCH] becomes true. At this point, the auction selects the buyer-optimal full bid $\eta_i$ for each seller $s_i$.

$$\eta_i = \arg\max_{\theta \in B_i^T}(\pi_b^T(\theta)). \qquad (8)$$

---

4. We drop the $t$ superscript in generic statements involving price and profit functions, understanding that all usage is with respect to the (currently) applicable prices.

In Phase B, $s_i$ may bid only on $\eta_i$. Sub-configuration prices are fixed at $p^T(\cdot)$ during this phase. The only adjustment is to $\Delta$, which is increased in every round by $\epsilon$. By (7), any increase of $\Delta$ decreases the current price of each of the configurations $\eta_i$. The auction terminates when at most one seller (if exactly one, designate it $s_{\hat{\imath}}$) is active. The allocation is determined according to four distinct cases:

1. All sellers drop out in Phase A (i.e., before rule [SWITCH] holds). The auction terminates with no allocation.

2. All active sellers drop out in the same round in Phase B. Of all the sellers that dropped in the last round, the auction selects a seller $s_i$ for which $u_b(\eta_i) - p^T(\eta_i)$ is maximal, and designates that seller as the winner $s_{\hat{\imath}}$. With a single winner, the appropriate case 3 or 4 is applied.

3. The auction terminates in Phase B with a final price above the buyer's valuation, $p^T(\eta_{\hat{\imath}}) - \Delta > u_b(\eta_{\hat{\imath}})$. It is still possible that there is exactly one seller (the winning seller) whose cost is below the buyer's valuation, in which case a trade with positive surplus is possible. Therefore, the auction offers the winner $s_{\hat{\imath}}$ an opportunity to supply $\eta_{\hat{\imath}}$ at price $u_b(\eta_{\hat{\imath}})$.

4. The auction terminates in Phase B with a final price $p^T(\eta_{\hat{\imath}}) - \Delta \leq u_b(\eta_{\hat{\imath}})$. This is the ideal situation, where the auction allocates the chosen configuration and seller at this resulting price.

---

Collect a reported valuation, $u_b(\cdot)$ from the buyer;
Set high initial prices, $p^1(\theta_r)$ on each sub-configuration $\theta_r$, and set $\Delta = 0$;
**while** *not [SWITCH], and at least one active seller* **do**
  | Collect sub-bids from sellers;
  | Compute $\mathcal{M}^t$;
  | Apply price change by [A];
**end**
Compute $\eta_i$;
**while** *more than one active seller* **do**
  | Increase $\Delta$ by $\epsilon$;
  | Collect bids on $(\eta_i, \Delta)$ from sellers;
**end**
Implement allocation and payment to winning seller;

**Algorithm 1**: GAI-based multiattribute auction.

---

The overall auction is described by high-level pseudocode in Algorithm 1. The role of Phase A is to guide the traders to their efficient configurations (MMP solutions), by reducing prices on configurations that are *chosen by at least one seller* but not preferred by the buyer. The price reduction makes such configurations slightly less attractive to the seller and slightly more attractive to the buyer. Phase B is a one-dimensional competition over the profit that remaining seller candidates can provide to the buyer. In the next section we formalize these notions, and prove that Phase A indeed converges and that Phase B selects a seller whose efficient configuration yields (approximately) the highest surplus. In Section 6.2 we discuss the computational tasks associated with the auction.

| | $I_1$ | | | | $I_2$ | | | |
|---|---|---|---|---|---|---|---|---|
| | $a^1b^1$ | $a^2b^1$ | $a^1b^2$ | $a^2b^2$ | $b^1c^1$ | $b^2c^1$ | $b^1c^2$ | $b^2c^2$ |
| $f_b$ | 65 | 50 | 55 | 70 | 50 | 85 | 60 | 75 |
| $f_1$ | 35 | 20 | **30** | 70 | 65 | **65** | 70 | 61 |
| $f_2$ | 35 | 20 | 25 | 25 | 55 | 110 | 70 | 95 |

Table 3: GAI utility functions for the example domain. $f_b$ represents the buyer's valuation, and $f_1$ and $f_2$ costs of sellers $s_1$ and $s_2$.

## 5. GAI Auction Example

We illustrate the auction with a simple three-attribute scenario, employing the two-element GAI structure $I_1 = \{a, b\}, I_2 = \{b, c\}$. Table 3 shows the GAI utilities for the buyer and the two sellers $s_1, s_2$. The efficient allocation is $(s_1, a^1b^2c^1)$: the buyer's valuation is $55 + 85 = 140$ and the cost of $s_1$ for this configuration (boldface in the table) is $30 + 65 = 95$, hence the surplus is 45. The maximal surplus of the second-best seller, $s_2$, is 25, achieved by $a^1b^1c^1$, $a^2b^1c^1$, and $a^2b^2c^2$. We set all initial prices over $I_1$ to 75, all initial prices over $I_2$ to 90, and $\epsilon = 8$, meaning that price reduction for sub-configurations ($\epsilon/g$) is 4.

For the sake of the example we assume that each seller bids in each round on the configuration that maximizes her profit (price minus cost), with respect to prices of the current round. In the next section we provide formal definitions and prove incentive properties for this strategy.

Table 4 shows the progress of Phase A. Initially all configuration have the same price (165), so sellers bid on their lowest-cost configuration—$a^2b^1c^1$ for both (with profit 80 to $s_1$ and 90 to $s_2$)—realized by sub-bids on $a^2b^1$ and $b^1c^1$. $\mathcal{M}^1$ contains the sub-configurations $a^2b^2$ and $b^2c^1$ of the highest value configuration $a^2b^2c^1$, which yields buyer profit of $-10$. As we show in the next section (Lemma 7), this maximum does not change throughout Phase A. Price is therefore decreased on $a^2b^1$ and $b^1c^1$. After the price change, the profit of $s_1$ for $a^2b^1c^1$ is 72, and because she has higher profit (74) on $a^1b^2c^2$ she bids on $a^1b^2$ and $b^2c^2$. Now (round 2) their prices go down, reducing the profit on $a^1b^2c^2$ to 66 and therefore in round 3 $s_1$ prefers $a^2b^1c^2$ (profit 67). Note that at this point the configuration $a^2b^2c^2$ yields profit of $-16$ to the buyer, which is within $\epsilon$ of the maximal buyer's profit (-10), hence $b^2c^2$ is marked to be in $\mathcal{M}^3$.

After the next price change, the configurations $a^1b^2c^1$ and $a^1b^2c^2$ both become optimal for $s_1$ (profit 66), and the sub-bids $a^1b^2$, $b^2c^1$ and $b^2c^2$ capture the two. These configurations stay optimal for another round (5), with profit 62. In round 5 the profit for configuration $a^1b^2c^1$ is $140 - 157 = -17$, which is within $\epsilon$ of maximizing the buyer's profit, therefore the sub-configuration $a^1b^2$ is added to $\mathcal{M}^5$. At this point $s_1$ has a full bid (in fact two full bids: $a^1b^2c^2$ and $a^1b^2c^1$) in $M^5$, and she no longer changes her bids because the price of her optimal configurations does not decrease. Seller $s_2$ however sticks to $a^2b^1c^1$ during the first four rounds, switching to $a^1b^1c^1$ in round 5. It takes four more rounds for $s_2$ and $\mathcal{M}^t$ to converge ($\mathcal{M}^9 \cap B_2^9 = \{a^1b^1c^1\}$).

| t | $I_1$ | | | | $I_2$ | | | |
|---|---|---|---|---|---|---|---|---|
| | $a^1b^1$ | $a^2b^1$ | $a^1b^2$ | $a^2b^2$ | $b^1c^1$ | $b^2c^1$ | $b^1c^2$ | $b^2c^2$ |
| 1 | 75 | 75 | 75 | 75 | 90 | 90 | 90 | 90 |
| | | $s_1,s_2$ | | $*$ | $s_1,s_2$ | $*$ | | |
| 2 | 75 | 71 | 75 | 75 | 86 | 90 | 90 | 90 |
| | | $s_2$ | $s_1$ | $*$ | $s_2$ | $*$ | | $s_1$ |
| 3 | 75 | 67 | 71 | 75 | 82 | 90 | 90 | 86 |
| | | $s_1,s_2$ | | $*$ | $s_2$ | $*$ | $s_1$ | $*$ |
| 4 | 75 | 63 | 71 | 75 | 78 | 90 | 86 | 86 |
| | | $s_2$ | $s_1$ | $*$ | $s_2$ | $*,s_1$ | | $*,s_1$ |
| 5 | 75 | 59 | 67 | 75 | 74 | 90 | 86 | 86 |
| | $s_2$ | | $*,s_1$ | $*$ | $s_2$ | $*,s_1$ | | $*,s_1$ |
| 6 | 71 | 59 | 67 | 75 | 70 | 90 | 86 | 86 |
| | | $s_2$ | $*,s_1$ | $*$ | | $*,s_1$ | $s_2$ | $*,s_1$ |
| 7 | 71 | 55 | 67 | 75 | 70 | 90 | 82 | 86 |
| | $s_2$ | | $*,s_1$ | $*$ | $s_2$ | $*,s_1$ | | $*,s_1$ |
| 8 | 67 | 55 | 67 | 75 | 66 | 90 | 82 | 86 |
| | $*$ | $s_2$ | $*,s_1$ | $*$ | $*$ | $*,s_1$ | $s_2$ | $*,s_1$ |
| 9 | 67 | 51 | 67 | 75 | 66 | 90 | 78 | 86 |
| | $*,s_2$ | | $*,s_1$ | $*$ | $*,s_2$ | $*,s_1$ | | $*,s_1$ |

Table 4: Auction progression in Phase A. Sell bids and designation of $\mathcal{M}^t$ (using $*$) are shown below the price of each sub-configuration.

After round 9, the auction sets $\eta_1 = a^1b^2c^1$ (which yields more buyer profit than $a^1b^2c^2$) and $\eta_2 = a^1b^1c^1$. In the second phase, which starts at this point, the sellers compete on the amount of surplus they transfer to the buyer, whose profit consequently becomes positive. For the next round (10) $\Delta = 8$, increased by 8 for each subsequent round. Note that $p^9(a^1b^1c^1) = 133$, and $c_2(a^1b^1c^1) = 90$, therefore the profit of $s_2$ at this point is 43. In round 15, $\Delta = 48$ meaning $p^{15}(a^1b^1c^1) = 85$ and that causes $s_2$ to drop out because his profit becomes negative. This ends the auction, and sets the final allocation to $(s_1, a^1b^2c^1)$ and $p^T(a^1b^2c^1) = 157 - 48 = 109$. That leaves the buyer with a profit of 31 and $s_1$ with a profit of 14.

# 6. Analysis

We analyze the economic properties of the auction in Section 6.1, and address practical and computational issues in Section 6.2.

## 6.1 Economic Properties

We adopt the following assumptions for this discussion:

**A$_1$** The optimal (seller, configuration) pair provides non-negative surplus.

**A$_2$** $u_b(\cdot)$ is the real utility function of the buyer.

When the optimal solution to *MAP* (3) provides negative surplus and sellers do not bid below their cost, the auction terminates in Phase A, no trade occurs, and the auction is trivially efficient. Therefore Assumption $A_1$ does not cause loss of generality. $A_2$ can be interpreted as follows: given non-truthful buyer report, our efficiency results below apply to the face value of the buyer's report rather than to the true utility.

### 6.1.1 Properties of the buyer's profit function

For $\mu \subseteq \{1, \ldots, g\}$, we define the (partial) profit from a set of of sub-configurations $\theta_\mu$ corresponding to $\mu$ as

$$\pi_b(\theta_\mu) = \sum_{r \in \mu} (f_{b,r}(\theta_r) - p(\theta_r)).$$

The functions $f$ come from the GAI breakdown of $u_b$ as in (6).

**Lemma 4.** *For any $\mu$ and its complement $\bar{\mu}$,*

$$\pi_b(\theta) = \pi_b(\theta_\mu) + \pi_b(\theta_{\bar{\mu}})$$

*Proof.* From (6) and from the definition of $\pi_b(\theta_\mu)$ we get

$$\pi_b(\theta) = \sum_{r \in \mu} (f_{b,r}(\theta_r) - p(\theta_r)) + \sum_{r \in \bar{\mu}} (f_{b,r}(\theta_r) - p(\theta_r)) = \pi_b(\theta_\mu) + \pi_b(\theta_{\bar{\mu}}).$$

$\square$

In round 5 of the example in the previous section, the sub-configuration $a^1 b^2$ is placed in $\mathcal{M}$ because the configuration $a^1 b^2 c^1$ is within $\epsilon$ of the maximal buyer profit $-10$. Actually, at that point not only $a^1 b^2 c^1$ is added to $M^t$, but also $a^1 b^2 c^2$ whose buyer profit $(-23)$ is not within $\epsilon$ of the maximum. If $a^1 b^2 c^2$ is later selected as $\eta_i$ for some $s_i$ this could lead to additional efficiency loss, beyond $\epsilon$. The following lemma bounds this potential loss.

**Lemma 5.** *Let $\Psi$ be a set of configurations, all within $\epsilon$ of maximizing profit for a trader $\tau$ (buyer or seller) at given prices. Let $\Phi = \{\theta_r \mid \theta \in \Psi, r \in \{1, \ldots, g\}\}$. Then any consistent cover in $\Phi$ is within $g\epsilon$ of maximizing profit for $\tau$ under the same prices.*

In particular, if $\Psi$ includes only exactly optimal configurations, any consistent cover will be exactly optimal as well. The proof (in Appendix B.1) relies on our definition of the GAI decomposition as a tree decomposition, and uses the partial profit function defined above along with Lemma 4.

The bound above is tight, in that for any GAI tree and nontrivial domain we can construct an example set $\Psi$ as above in which there exists a consistent cover whose utility is exactly $g\epsilon$ below the maximal.

As a result we get the following corollary.

**Corollary 6.**

$$\forall \theta \in M^t. \ \pi_b^t(\theta) \geq \max_{\theta' \in \Theta} \pi_b^t(\theta') - g\epsilon$$

*Proof.* Apply Lemma 5 for $\pi_b^t$: define $\Psi$ as the set of configurations within $\epsilon$ of $\max_{\theta' \in \Theta} \pi_b^t(\theta')$. $\mathcal{M}^t$, by its definition, serves as $\Phi$ in the lemma. $M^t$ is then exactly the set of consistent covers over $\Phi$, and hence each $\theta \in M^t$ must be within $g\epsilon$ of the optimum $\max_{\theta' \in \Theta} \pi_b^t(\theta')$. □

We now show that, as noted in the example, the maximal profit of the buyer does not change during Phase A.

**Lemma 7.** $\max_{\theta \in \Theta} \pi_b^t(\theta) = \max_{\theta \in \Theta} \pi_b^1(\theta)$ *for any round $t$ of Phase A.*

*Proof.* Assume there exists some $\theta'$ for which $\pi_b^{t+1}(\theta') > \pi_b^t(\theta')$. Then necessarily $p^{t+1}(\theta') = p^t(\theta') - \delta$ for some $\delta > 0$. The only price change is by Rule [A], meaning that some $w \leq g$ sub-configurations of $\theta'$ are not in $\mathcal{M}^t$, and $\delta = \frac{w\epsilon}{g}$. In that case, by definition of $\mathcal{M}^t$,

$$\pi_b^t(\theta') < \max_{\theta \in \Theta} \pi_b^t(\theta) - \epsilon.$$

Therefore,

$$\pi_b^{t+1}(\theta') = \pi^t(\theta') + \delta = \pi^t(\theta') + \frac{w\epsilon}{g} \leq \pi^t(\theta') + \frac{g\epsilon}{g} < \max_{\theta \in \Theta} \pi_b^t(\theta).$$

This is true for any $\theta'$ whose profit improves, therefore $\max_{\theta \in \Theta} \pi_b^t(\theta)$ does not change during Phase A, and hence equals its value in round 1. ■

#### 6.1.2 STRAIGHTFORWARD BIDDING SELLERS

We now turn our attention to the sellers' behavior. We first define the profit function of seller $s_i$ by $\pi_i^t(\theta) = p^t(\theta) - c_i(\theta)$.

**Definition 14 (Straightforward Bidder).** A seller is called a *straightforward bidder* (SB) if at each round $t$ she bids on $\mathcal{B}_i^t$ as follows: if $\max_{\theta \in \Theta} \pi_i^t(\theta) < 0$, then $\mathcal{B}_i^t = \emptyset$. Otherwise select $b_i^t \in \arg\max_{\theta \in \Theta} \pi_i^t(\theta)$, and set

$$\mathcal{B}_i^t = \{\theta_r \mid \theta \in b_i^t, r \in \{1, \dots, g\}\}.$$

Intuitively, SB sellers follow a *myopic best response* strategy, optimizing profit with respect to current prices. This approach was termed "straightforward" by Milgrom (2000) in the sense that agents bid myopically, rather than strategically anticipating subsequent price responses.

SB sellers can choose *any* optimal configuration to bid on; none of the results proved below is affected by this choice. It is also important to note that SB sellers find their optimal full *configuration* $b_i^t$, rather than optimize each GAI element separately. The configuration $b_i^t$ is translated to its set of sub-configurations $\mathcal{B}_i^t$. In order to calculate $b_i^t$, seller $s_i$ needs to find the optimum of her current profit function. In Section 6.2 we show that this optimization problem is tractable under the assumption that $u_i(\cdot)$, too, has a compact GAI structure.

The following is an immediate corollary of the definition of SB.

**Corollary 8.** *For SB seller $s_i$,*

$$\forall t, \forall \theta \in B_i^t. \ \pi_i^t(\theta) = \max_{\theta' \in \Theta} \pi_i^t(\theta').$$

In general, sellers' preference structure may not coincide with the auction's price structure. Nevertheless, Corollary 8 holds by definition of SB, because $B_i^t$ (defined in Section 4.2) contains a single configuration which is the submitted bid $b_i^t$. Alternatively, the definition of SB can be modified, so sellers with GAI preferences consistent with the auction's price structure can bid on multiple optimal configurations (if such exist). If sellers bid on multiple configurations, this can speed up convergence. In that case $b_i^t$ denotes a set of submitted configurations, $\mathcal{B}_i^t$ denotes the respective collection of sub-configurations, and $B_i^t$ is the set of consistent covers over $\mathcal{B}_i^t$. Lemma 5 (with $\epsilon = 0$) entails that Corollary 8 still holds. However, for simplicity of the analysis we retain Definition 14.

### 6.1.3 Efficiency given SB

Lemma 9 states that through the price system and price change rules, Phase A leads the buyer and each of the sellers to their mutually efficient configuration. Formally, we are interested in maximizing the function $\sigma_i : \Theta \to \Re$, which represents the surplus $u_b(\cdot) - c_i(\cdot)$. For any prices $p^t$,

$$\sigma_i(\theta) = \pi_b^t(\theta) + \pi_i^t(\theta).$$

**Lemma 9.** *For SB seller $s_i$, $\eta_i$ is $g\epsilon$-efficient:*

$$\sigma_i(\eta_i) \geq \max_{\theta \in \Theta} \sigma_i(\theta) - g\epsilon.$$

*Proof.* Configuration $\eta_i$ is chosen to maximize the buyer's profit out of $B_i^t$ at the end of Phase A. Because $B_i^t \cap M^t \neq \emptyset$, a configuration $\eta_i \in M^t$ is available in $B_i^t$, hence one must be chosen to maximize buyer's utility. For any $\tilde{\theta}$, and for *any* $\eta_i \in B_i^t$, we get from Corollary 8,

$$\pi_i^T(\eta_i) \geq \pi_i^T(\tilde{\theta}),$$

and from Corollary 6, we get for *any* $\eta_i \in M^t$,

$$\pi_b^T(\eta_i) \geq \pi_b^T(\tilde{\theta}) - g\epsilon.$$

Because $\eta_i \in B_i^t \cap M^t$ we can add up the two inequalities and get $\sigma_i(\eta_i) \geq \sigma_i(\tilde{\theta}) - g\epsilon$, which is the desired result. $\square$

Based on Phase B's simple role as a single-dimensional bidding competition over the discount, we next assert that the overall result is efficient under SB, which in turn (Section 6.1.4) proves to be an approximately ex-post equilibrium strategy in the two phases.

**Theorem 10.** *Given a truthful buyer and SB sellers, the surplus of the final allocation is within $(g + 1)\epsilon$ of the maximal surplus.*

*Proof Sketch:* we first establish that the auction must reach Phase B. To do that, we show that in each round of Phase A, a price of at least one sub-configuration is reduced, whereas by Lemma 7, $\max_{\theta \in \Theta} \pi_b^t(\theta)$ does not change. The latter enforces a lower bound on how far prices can be reduced within Phase A, hence Phase A must terminate. Because initial prices are above the buyer's valuation, a seller whose surplus (MMP solution) is positive cannot drop during that phase, so using Assumption $A_1$ we show that the only way for Phase A to terminate is by reaching condition [SWITCH]. Next, we show that for any two sellers, the surplus of the first to drop from the auction cannot be significantly higher than that of the one who stayed longer. This ensures that the winning seller is the efficient one, or one whose MMP surplus is almost maximal, and from Lemma 9 the auction must obtain (almost) all of that surplus. The full proof is given in Appendix B.2.

The bound guaranteed by Theorem 10 is a worst-case bound, and as shown experimentally in the following sections the auction typically achieves efficiency closer to the optimum. In the example of Section 5, the difference in the efficiencies of the two sellers is lower than the potential efficiency loss (as $(g + 1)\epsilon = 24$). However, for that instance it is still guaranteed that $s_1$ wins, either with the efficient allocation, or with $a^1 b^2 c^2$ which provides a surplus of 39. The reason is that these are the only two configurations of $s_1$ with surplus within $g\epsilon = 16$ of the solution to $MMP(b, s_1)$, hence by Lemma 9 one of them must be chosen as $\eta_1$. Both of these configurations provide more than $\epsilon$ surplus over $s_2$'s most efficient configuration, and this is sufficient in order to win in Phase B.

The bound of Theorem 10 can be improved when the CDI map contains disconnected components. For example, when a fully additive decomposition (as assumed in previous literature) does exist, the CDI map contains a disconnected component for each attribute. To take advantage of this disconnectedness we create a separate tree decomposition for each disconnected components. The definition of $\mathcal{M}$ has to be adapted to apportion $\epsilon$ proportionally across the disconnected trees. Formally, we redefine $\mathcal{M}^t$ as follows.

**Definition 15** (**Buyer's Preferred Set**). Let $G$ be comprised of trees $G_1, \ldots, G_h$. Let $\theta_j$ denote the projection of a configuration $\theta$ on the tree $G_j$, and $g_j$ the number of GAI elements in $G_j$. Similarly, $\Theta_j$ denotes the projection of $\Theta$ on $G_j$. Define

$$\mathcal{M}_j^t = \{\theta_r \mid \pi_b^t(\theta_j) \geq \max_{\theta_j' \in \Theta_j} \pi_b^t(\theta_j') - g_j \frac{\epsilon}{g}, r \in G_j\}.$$

The *buyer's preferred set* is given by $\mathcal{M}^t = \bigcup_{j=1}^h \mathcal{M}_j^t$.

Let $e_j = g_j - 1$ denote the number of edges in $G_j$. We define a *connectivity parameter*, $e = \max_{j=1,\ldots,h} e_j$. It turns out that $e + 1$ can replace $g$ in the approximation results. The first step is to replace Corollary 6 with this tighter bound on the optimality of configurations in $M^t$.

**Corollary 11.**
$$\forall \theta \in M^t. \ \pi_b^t(\theta) \geq \max_{\theta' \in \Theta} \pi_b^t(\theta') - (e + 1)\epsilon$$

*Proof.* We apply Lemma 5 for each $G_j$, but with $g_j \frac{\epsilon}{g}$ instead of $\epsilon$, hence any consistent cover over $\mathcal{M}_j^t$ is within $g_j \frac{\epsilon}{g} g_j$ of $\max_{\theta_j' \in \Theta_j} \pi_b^t(\theta_j')$. From Lemma 4, we get that any consistent cover over $\mathcal{M}^t$ (meaning any configuration in $M^t$) is within $\sum_{r=1}^h g_j \frac{\epsilon}{g} g_j$ of $\max_{\theta' \in \Theta} \pi_b^t(\theta')$. As $e + 1 = \max_{j=1,\ldots,h} g_j$, this is bounded by $\frac{\epsilon}{g} \sum_{r=1}^h g_j(e + 1) = \epsilon(e + 1)$. $\qquad\square$

We can now obtain a tighter efficiency result.

**Theorem 12.** *Given a truthful buyer and SB sellers, the surplus of the final allocation is within $(e + 2)\epsilon$ of the maximal surplus.*

In the fully additive case this loss of efficiency reduces to $2\epsilon$. On the other extreme, if the CDI map is connected then $e + 1 = g$, reducing Theorem 12 to Theorem 10. If we do not assume any preference structure for the buyer, meaning that the CDI map is *fully* connected, then $e = 0$ and the efficiency loss is again proportional to $\epsilon$.

### 6.1.4 Sellers' incentives to use SB

Following Parkes and Kalagnanam (2005), we relate our auction to the Vickrey-Clarke-Groves (VCG) mechanism to establish incentive properties for the sellers. In the one-sided multiattribute VCG auction, the buyer reports valuation $u_b$, the sellers report cost functions $\hat{c}_i$, and the buyer pays the sell-side VCG payment to the winning seller.

**Definition 16 (Sell-Side VCG Payment).** Let $(\theta^*, i^*)$ be an optimal solution to *MAP*. Let $(\tilde{\theta}, \tilde{i})$ be the best solution to *MAP* when $i^*$ does not participate. The *sell-side VCG payment* is

$$VCG(u_b, \hat{c}_i) = u_b(\theta^*) - \max(0, u_b(\tilde{\theta}) - \hat{c}_{\tilde{i}}(\tilde{\theta})).$$

It is well known that truthful bidding is a dominant strategy for sellers in the one-sided VCG auction. Parkes and Kalagnanam (2005) showed that the maximal regret for buyers from bidding truthfully in this mechanism is $u_b(\theta^*) - c_{i^*}(\theta^*) - (u_b(\tilde{\theta}) - \hat{c}_{\tilde{i}}(\tilde{\theta}))$, that is, the *marginal product* of the efficient seller.

As typical for iterative auctions, the VCG outcome is not exactly achieved, but the deviation is bounded by the minimal price change.

**Definition 17 ($\delta$-VCG Payment).** A *sell-side $\delta$-VCG payment* for *MAP* is a payment $p$ such that

$$VCG(u_b, \hat{c}_i) - \delta \leq p \leq VCG(u_b, \hat{c}_i) + \delta.$$

When payment is guaranteed to be $\delta$-VCG, sellers can affect their payment only within that range, hence their gain from falsely reporting cost is bounded by $2\delta$.

**Lemma 13.** *When sellers are SB, the GAI auction payment is sell-side $(e + 2)\epsilon$-VCG.*

In the example of Section 5, the profit of the winner (14) is less than $\epsilon$ below his VCG profit 20. The proof (in Appendix B.4) also covers Case 3 in the allocation options of Section 4.2, in which we force the payment to equal $u_b(\eta_{\tilde{i}})$.

We are now ready for our final result of this section, showing that the approximately efficient outcome guaranteed by Theorem 12 is achieved in an (approximate) ex-post Nash equilibrium.

**Theorem 14.** *SB is a $(3e + 5)\epsilon$ ex-post Nash equilibrium for sellers in the GAI auction. That is, sellers cannot gain more than $(3e + 5)\epsilon$ by deviating from SB, given that other sellers follow SB.*

In order to exploit even this bounded potential gain, sellers need to know, for a given configuration in $M^t$, whether it was explicitly selected as approximately optimal for the buyer, or it is a combination of sub-configurations from approximately optimal configurations. It seems highly unlikely for sellers to have such information. They are more likely to lose if they do not bid on their myopically optimal configurations.

## 6.2 Computation and Complexity

The advantage of GAI auctions over an additive auction such as AD (Parkes & Kalagnanam, 2005) is in economic efficiency: by accommodating expressive bidding, the efficiency results are with respect to a more accurate utility function. In contrast, the key advantage with respect to an auction that does not employ preference structures, such as auction NLD (Parkes & Kalagnanam, 2005), is in computational efficiency. The property we show in this section is that all computations are exponential only in the size of the largest GAI element, rather than in $|A|$. In particular, the size of the price space the auction maintains is equal to the total number of sub-configurations. This number is exponential in the treewidth (plus one) of the original CDI map.[5] To ensure computational tractability, one can define *a priori* a constant $C$, and force the treewidth of the CDI map to be bounded by $C$ by ignoring some of the interdependencies. This is still much better than using an additive representation that ignores all interdependencies. The constant represents a tradeoff between economic and computational efficiency; a larger $C$ supports a more accurate preference representation, but the GAI elements may be larger.

For the purpose of computational analysis, let $\mathcal{I} = \bigcup_{r=1}^{g} \prod_{a_j \in I_r} D(a_j)$, that is the collection of all sub-configurations. Since $M^t$ grows monotonically with $t$, naïve generation of the best outcomes sequentially might end up enumerating significant portions of the domain. Fortunately, this enumeration can be avoided, and the complexity of this computation (as well as the optimization performed by the seller) grows only with $|\mathcal{I}|$, that is, no computation depends on the size of the exponential domain.

**Theorem 15.** *The computation of $\mathcal{M}^t$ can be performed in time $O(g|\mathcal{I}|^2)$. Moreover, the total time spent on this task throughout the auction is $O(g|\mathcal{I}|(|\mathcal{I}| + T))$.*

We obtain a bound on $T$, the number of rounds of Phase A, by comparing the sum of prices of all sub-configurations in rounds 1 and $T$.

**Theorem 16.** *The number of rounds required by the auction is bounded by*

$$T \leq \sum_{\theta_r \in \mathcal{I}} p^1(\theta_r)\frac{g}{\epsilon}.$$

*Proof.* Let $\Sigma^i = \sum_{\theta_r \in \mathcal{I}} p^i(\theta_r)$ (the sum of prices of all sub-configurations in round $i$). Assume that $\Sigma^i < 0$ for some $1 \leq i \leq T$. Then because $u_b(\cdot) \geq 0$, there must exist $\theta \in \Theta$ for which $\pi_b^i(\theta) > 0$. Because we chose initial prices such that for all $\theta \in \Theta$, $\pi_b^1(\theta) < 0$, this contradicts Lemma 7. Therefore, $\Sigma^T \geq 0$, hence the sum of prices cannot be reduced by more than $\Sigma^1 = \sum_{\theta_r \in I} p^1(\theta_r)$ throughout the auction. Also, in each round at least one price is reduced by $\frac{g}{\epsilon}$. This leads to the required result. $\qquad\square$

This bound is rather loose—its purpose is to ensure that the number of rounds does not depend on the size of the non-factored domain. It depends on the number of sub-configurations, and on the result of dividing the initial price by the minimum price decrement. Usually Phase A converges much faster. Let the initial negative profit chosen by the auctioneer be $m = \max_{\theta \in \Theta} \pi_b^1(\theta)$. In the worst case, Phase A needs to run until $\forall \theta \in \Theta$. $\pi_b(\theta) = m$. This happens for example when $\forall \theta_r \in I$. $p^t(\theta_r) = f_{b,r}(\theta_r) + \frac{m}{g}$. That implies that the closer the initial prices reflect the buyer's valuation, the faster Phase A converges. One extreme choice is to set $p^1(\theta_r) = f_{b,r}(\theta_r) + \frac{m}{g}$. That would make Phase A redundant, at the cost of full initial revelation of the buyer's valuation (Section 2.3). Between this option and the other extreme, which is $\forall \alpha, \hat{\alpha} \in I$. $p^1(\alpha) = p^1(\hat{\alpha})$, the auctioneer has a range of choices to determine the right tradeoff between convergence time and information revelation. In the example of Section 5, the choice of a lower initial price for the domain of $I_1$ provides some speedup by revealing a harmless amount of information. In our simulations below, we also set constant initial prices within each GAI element.

---

5. The use of the term treewidth is subject to using an optimal tree decomposition.

Furthermore, many domains have natural dependencies that are mutual to traders, in which case the price structure used by the auction may also accommodate sellers' preference structures. If so, sellers can bid on multiple equally profitable configurations in each round, thus speeding up convergence, as discussed above in Section 6.1.

We also consider computational complexity of the SB strategy for sellers.

**Theorem 17.** *Let $\rho_b$ denote the treewidth of the CDI map of $u_b(\cdot)$, and let $\rho_i$ denote the treewidth of the CDI map of $u_i(\cdot)$. The optimization of $u_i(\cdot) - p(\cdot)$ takes time exponential in $\rho_b + \rho_i$ in the worst case.*

*Proof.* Consider the graph $G$ which includes the union of the edges of the two CDI maps. The treewidth of $G$ is $\rho_b + \rho_i$ in the worst case. By definition, the price function $p(\cdot)$ is decomposed according to $u_b(\cdot)$, hence $u_i(\cdot) - p(\cdot)$ decomposes according to the additive GAI factors of $u_i(\cdot)$ and $u_b(\cdot)$. Therefore, for any pair of attributes $x$ and $y$ which have a mutual factor in $u_i(\cdot) - p(\cdot)$, there is an edge $x, y$ in $G$. It is well known that the complexity of combinatorial optimization is exponential only in the treewidth of such graph—for example, using cost networks (Dechter, 1997). □

Of potential concern may be the communication cost associated with the descending auction style. The sellers need to send their bids over and over again at each round. A simple change can be made to avoid much of the redundant communication: the auction can retain sub-bids from previous rounds on sub-configurations whose price did not change. Because combinations of sub-bids from different rounds can yield suboptimal configurations, each sub-bid should be tagged with the number of the latest round in which it was submitted, and only consistent combinations *from the same round* are considered to be full bids. With this implementation sellers need not resubmit their bid until a price of at least one of its sub-configurations has changed.

To summarize, GAI auctions are shown to perform well on the criteria mentioned in Section 2.3: they achieve approximate efficiency given reasonable incentive properties, they are expressive enough to accommodate preferences with interdependencies among attributes, they are tractable when the maximal size of GAI clusters is reasonably bounded, and they do not require full revelation of utility. Performance on this last criterion is quantified in the experimental part of the paper.

## 7. Experimental Design

The main idea behind GAI auctions is to improve efficiency over auctions that assume additivity, when preferences are not additive. However, (given a fixed $\epsilon$) the theoretical efficiency guarantee of GAI auctions depends on $e$, the connectivity parameter of the GAI tree. This suggests a tradeoff: complex models more accurately represent true utility, but can increase approximation error due to higher connectivity. An obvious question is whether more accurate preference modeling is indeed more efficient, and in particular, whether GAI auctions are more efficient than additive auctions, given that the preferences are not additive. To address the question experimentally, we assume that the buyer's preferences have some GAI structure, and compare the performance of GAI auctions that model this structure with the performance of auctions that are restricted to an additive representation. For the latter, we use an instance of GAI auction in which the pricing structure is additive, and name it the *additive approximating* auction (AP). This auction is similar in principle to auction AD (Parkes & Kalagnanam, 2005).[6] To the best of our knowledge, AD is the only proposed instance of a surplus-maximizing multiattribute auction based on additive preferences, besides those that require full revelation of the buyer's utility. In all of the experiments, sellers employ SB strategies.

In Section 7.1 we describe how random GAI utilities are drawn, and in Section 7.2 we extend the scheme to generate GAI utility functions that exhibit additional structure. In Section 7.3 we show how we obtain an additive approximation of these random functions, allowing us to simulate auction AD. The results of the simulations are presented in Section 8.

## 7.1 GAI Random Utility

We performed simulations using randomly generated utility functions representing the buyer's value function and sellers' cost functions. Our random utility generation procedure follows the utility elicitation procedure suggested by Braziunas and Boutilier (2005), and uses a two-step process: first we create *local* utility functions over each GAI element, normalized to the range $[0, 1]$. Next, we draw scaling constants that represent the relative weight of each local function in the overall utility.

More formally, let $\bar{u}_r(I_r) = u([I_r])$ denote a *local* utility function over $I_r$, each normalized to $[0, 1]$. Next, let $\bar{f}_r(I_r)$ be defined according to the GAI functional form of Eq. (6), with $u([I_r])$ replaced with $\bar{u}_r(I_r)$, hence

$$\bar{f}_1 = \bar{u}_1(I_1), \text{ and}$$

$$\text{for } r = 2, \ldots, g, \quad \bar{f}_r = \bar{u}_r(I_r) + \sum_{j=1}^{r-1} (-1)^j \sum_{1 \leq i_1 < \cdots < i_j < r} \bar{u}_r([\bigcap_{s=1}^{j} I_{i_s} \cap I_r]). \quad (9)$$

Braziunas and Boutilier (2005) show that for GAI-structured utility, there exist scaling constants $\lambda_r \in [0, 1]$ such that

$$\bar{u}(A) = \sum_{r=1}^{g} \lambda_r \bar{f}_r(I_r). \quad (10)$$

---

6. Both auctions employ an additive price space that drives bidders to their efficient configurations. AD is efficient up to $\epsilon$ when the buyer and all the sellers have additive preferences. GAI auctions are $\epsilon$-efficient given additive buyer's preferences, and make no assumption regarding sellers' preference. There are some more structural differences: (i) AD employs more complicated price change rules, in order to allow sellers to ignore some of the attributes, (ii) discounts can be used in any stage of AD, and the auction selects a provisional winner at each iteration.

We refer to the functions $\bar{u}_r(I_r)$ as *subutility functions*. Note that values of the form $\bar{u}_r([I_{i_r} \cap I_{i_{r'}}])$ are drawn only once and used in both $\bar{u}_r(I_r)$ and $\bar{u}_r(I_{i_{r'}})$. This representation lets us draw random GAI functions, for a given GAI tree structure, using the following steps:

1. Draw random subutility functions $\bar{u}_r(I_r), r = 1, \ldots, g$ in the range $[0,1]$.

2. Compute $\bar{f}_r(\cdot), r = 1, \ldots, g$ using (9).

3. Draw random scaling constants $\lambda_r$, such that $\sum_{r=1}^{g} \lambda_r = 1$, and compute $\bar{u}(A)$ by (10).

The scaling constants represent the importance the decision maker accords to corresponding GAI elements in the overall decision. The procedure results in utilities that are normalized in $[0, 1]$. Finally, for each particular trader we draw mean $\mu$ and variance $\sigma$, and scale $\bar{u}(\cdot)$ to the range $[\mu - \sigma, \mu + \sigma]$, resulting in the utility functions $u_b(\cdot)$ and $u_i(\cdot)$ for $i = 1, \ldots, m$.

## 7.2 Structured Subutility

A subutility function in the model above may represent any valuation over the subspace. In practice we may often find additional structure within each GAI element. We introduce two structures which we consider most typical and generally applicable, and we use them for the simulations, along with completely random local functions.

As we argue in Section 2.1, typical purchase and sale decisions exhibit FOPI (first order preferential independence), meaning that most or all single attributes have a natural ordering of quality. For example, hard-drive buyers always prefer more memory, higher RPM, longer warranty, and so on. To implement FOPI, we let the integer values of each attribute represent its quality. For example, if $a$ belongs to some GAI element $I_r = \{a, b\}$, we make sure that $\bar{u}_r(a_i, b') \geq \bar{u}_r(a_j, b')$ for any $a_i > a_j$, $a_i, a_j \in D(a)$, and any $b' \in D(b)$. This must of course hold for any attribute $a$ that is FOPI, and any GAI element $I_r$ that includes $a$. We enforce the condition after all the values for that GAI element have been drawn, through a special-purpose sorting procedure, applied between steps 1 and 2 above.

The FOPI condition makes the random utility function more realistic, and in particular more appropriate to the target application. Once attributes exhibit FOPI, the dependencies among different attributes are likely to be framed as complements or substitutes. These concepts are known primarily in the context of combinatorial preferences, that is, preferences over combinations of distinct items. In the multiattribute framework, two attributes are complements if an improvement in the value of both is worth more than the sum of the same improvement in each separately. Two attributes are substitutes if it is the other way around. These concepts are meaningful only with respect to attributes that are FOPI, otherwise the notion of improvement is conditional on the value of other attributes.

**Definition 18 (Complements and Substitutes).** Let $u(\cdot)$ be a measurable value function over $S'$. Let $a, b \in S'$, and $Z = S' \setminus \{a, b\}$, and assume that $a$ and $b$ are each FOPI of the rest of the attributes. Attributes $a$ and $b$ are called *complements* if for any $a^i > a^{\hat{i}}$ $(a^i, a^{\hat{i}} \in D(a))$ and any $b^j > b^{\hat{j}}$ $(b^j, b^{\hat{j}} \in D(b))$, and any $Z' \in D(Z)$,

$$u(a^i, b^j, Z') - u(a^{\hat{i}}, b^{\hat{j}}, Z') > u(a^i, b^{\hat{j}}, Z') - u(a^{\hat{i}}, b^{\hat{j}}, Z') + u(a^{\hat{i}}, b^j, Z') - u(a^{\hat{i}}, b^{\hat{j}}, Z').$$

Attributes $a$ and $b$ are substitutes if the inequality sign is (always) reversed.

This relationship between attributes is ruled out under an additive utility function, but admitted under a weaker independence condition, called *mutual utility independence* (MUI) (Keeney & Raiffa, 1976), which implies that the utility function can be either multiplicative or additive. If it is multiplicative, the utility function can be represented by $n$ single-dimensional functions, $n$ scaling constants, and a single parameter $k$ (the *MUI-factor*) that controls the strength of complementarity ($k > 0$) or substitutivity ($k < 0$) between all pairs of attributes within a GAI element (for $k = 0$ the set of attributes is additive).[7] For experimental purposes, we assume that each attribute cluster (GAI element) exhibits MUI, and that the value of $k$ is the same for all.

In an elicitation procedure, one would normally extract the MUI scaling constants from a user, and then compute $k$ (Keeney & Raiffa, 1976). For our purposes, we first determine $k$ according to the relationship we wish to impose on the attributes, and then draw MUI scaling constants that are consistent with this value. More explicitly, we draw random scaling constants, and then iteratively modify all the constants, until a set of constants is found that is consistent with $k$. The next step is to compute $\bar{u}_r(I_r)$ according to the MUI formula (Keeney & Raiffa, 1976). The $\bar{u}_r(I_r)$ (for all $r$) are in the range $[0, 1]$, hence at this point we can proceed with steps 2 and 3 above. Note that in this procedure several distinct sets of scaling constants are used: the $g$ constants used in step 3 scale the different GAI elements, whereas the MUI constants, per GAI element, scale the attributes *within* the element.

## 7.3 Additive Approximation

Another issue for experiment design is how the additive auction (AP) behaves in the face of non-additive buyer preferences, specifically how would it select the approximately buyer-preferred sets of configurations. The approach we took is to come up with an additive function that approximates the buyer's true utility function, and use it throughout the auction. We are not aware of a better strategy, but do not rule out the possibility that one exists.

A natural approach to generate a linear approximation $\sum_i \hat{f}_i(\cdot)$ for an arbitrary function $u_b(\cdot)$ is to use linear regression. We define an indicator variable $x_{i_j}$ for every $a_{i_j} \in D(a_i)$, and consider any value of an assignment as a data point. For example, the assignment $a_{1_{j(1)}}, \ldots, a_{m_{j(m)}}$ creates the following data point:

$$\sum_{i=1}^{m} \sum_{a_{i_j} \in D(a_i)} c_{i_j} x_{i_j} = u(a_{1_{j(1)}}, \ldots, a_{m_{j(m)}}),$$

in which the value of the variable $x_{i_j}$ is 1 if $j = j(i)$ and 0 otherwise. The coefficients $c_{i_j}$ result from the regression and represent the values to be used as $\hat{f}_i(a_{i_j})$.

When the problem includes many attributes, it is not possible to consider all the points in $\Theta$. Under the assumption that a compact GAI representation exists, it is sensible to expect that we could use fewer data points for the regression. We indeed found that a small random sample from the joint utility yields an approximation as effective as one based on all the points. More precisely, for the largest domain we tested (25 attributes, each with domain of size 4) we found that the efficiency of AP does not improve when increasing the number of sampled points beyond 200. We show a chart supporting this claim in Appendix E. Our experiments use 300 points for all instances.

---

7. We formalize this notion in Appendix D.

This method of comparison probably overestimates the quality of an additive approximation. In general, we would not have the true utility function explicitly when we generate the approximation. Extraction or elicitation of the utility function is usually the most serious bottleneck of a multiattribute mechanism. Therefore, the major reason to use an additive approximation is to reduce the burden of elicitation. Hence in practice we would try to obtain the additive function directly, rather than obtain the full utility and then approximate it. The result of such process is somewhat unpredictable, because the elicitation queries may not be coherent: if willingness to pay for $a_1$ depends on the value of $b$, then what is the willingness to pay for $a_1$ when we do not know $b$? We therefore consider our experimental generation method biased in favor of the additive approximation.

## 8. Simulation Results

We provide detailed results of our simulation study. Section 8.1 provides and analyses economic efficiency results. Section 8.2 covers the computational study, and results regarding revelation of private information are provided in Section 8.3.

### 8.1 Efficiency and GAI Structure

We measure efficiency in terms of percentage of the MAP solution, which is the surplus achieved by the optimal seller-configuration pair. To evaluate the effect of preference modeling on efficiency, we vary structural parameters of the buyer's GAI preferences: the connectivity factor $e$, and the size $\xi$ of the largest GAI element. Performance depends on many additional factors, such as the size of attribute domains, number of sellers, amount of price decrement ($\epsilon$), and the distribution from which utility functions are drawn. To isolate the primary structural parameters, we first tested how efficiency varies according to the choices of these side factors, for several fixed GAI structures with fully random subutility functions. As a result of these tests, we picked the following parameter values for the rest of the simulations: all valuations are drawn from a uniform distribution, with buyer mean set at 500. A mean for each seller is drawn uniformly from $[500, 700]$. The variance is set at 200 for all traders. We use the same domain size of 3 or 4 for all attributes, and number of sellers $m = 5$. Further explanation of the process leading to these choices is provided in the full report (Engel, 2008).

In the following experiment we used a roughly fixed GAI structure, with $g = 6$ elements and $e = 5$, (that is, the GAI structure is a tree, not a forest), and $\epsilon = 24$ (meaning reduction of $\delta = 4$ per sub-configuration). We vary the number of attributes by varying the size of each element. Figure 2a shows the efficiency obtained with respect to $\xi$, the size of the largest GAI element. As expected, the size of the GAI elements has negligible, or no effect on the efficiency of GAI auctions. It has a dramatic effect on the efficiency of AP. When $\xi = 1$, the decomposition is in fact additive and hence AP performs optimally. The performance then deteriorates as $\xi$ increases.

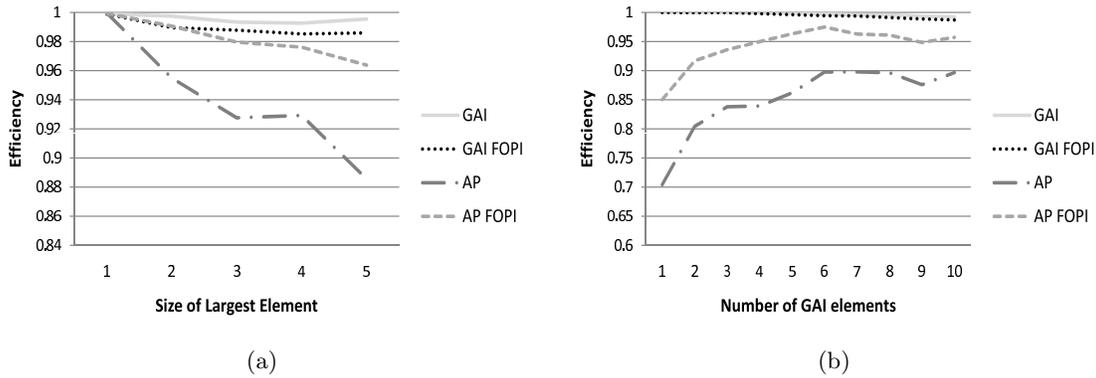

(a)                                    (b)

Figure 2: Efficiency as a function of: (a) the size of largest GAI element ($\xi$), given $e = 5$, (b) the number of GAI elements ($e + 1$), given $\xi = 5$.

We performed the same test when using utility in which all attributes are FOPI. With the FOPI restriction, the additive approximation is much more efficient relative to unconstrained random utility. When FOPI applies to a strict subset of the attributes, we would expect the efficiency of AP to fall somewhere between its efficiency under FOPI and the unrestricted case. Somewhat surprisingly, imposing FOPI renders the GAI auctions slightly less efficient. Nevertheless, the additive approximation achieves lower efficiency compared to the accurate preference modeling, with differences that pass the statistical significance test ($P < 0.01$), for $\xi \geq 3$. Further, note that the performance of GAI auctions can always be improved using a smaller value of $\epsilon$ and $\delta = \frac{\epsilon}{9}$, whereas this hardly improves performance of AP. With $\delta = 2$, a statistically significant difference (with the same confidence level) is already detected for $\xi \geq 2$. We used $\delta = 2$ hereafter.

The next experiment (Figure 2b) measures efficiency as a function of $e$, for a given fixed $\xi$. We assume connected GAI trees, so $e$ is the number of GAI elements minus one. We tested structures with $e$ varying from 1 to 10, all elements of size 3 to 5, and $\xi = 5$ for all the structures.[8] On a single element, the GAI auction is similar to NLD (Parkes & Kalagnanam, 2005), which is an auction that assigns a price to every point in the joint domain. Here $e = 0$, hence the efficiency of GAI is close to perfect. This structure is on the other extreme compared to an additive representation, and indeed the performance of AP is particularly inferior (only 70% efficient).

With more GAI elements, the efficiency of GAI auctions declines at a very slow pace. The theoretical potential error $(e + 2)\epsilon$, is mostly a result of efficiency loss of $\eta_i$ for the winning seller, based on Lemma 5. Such efficiency loss may occur only if each sub-configuration in $\eta_i$ belongs to a configuration that yields the lowest profit allowed in the buyer-preferred set—a particularly rare case. In practice, the loss is closer to $e\delta$, which is a much smaller error.

The performance of AP improves as the number of elements grows while their maximal and average sizes are fixed. Intuitively, changing the structure that way takes it closer to an additive representation. Under FOPI, we see a similar phenomenon as before. However, the difference between GAI FOPI and AP FOPI, even for ten elements, is substantial and statistically significant.

___________

8. We did not find the particular tree structure to be influential on the results; the final structure used in the reported results has a maximum of three children per node.

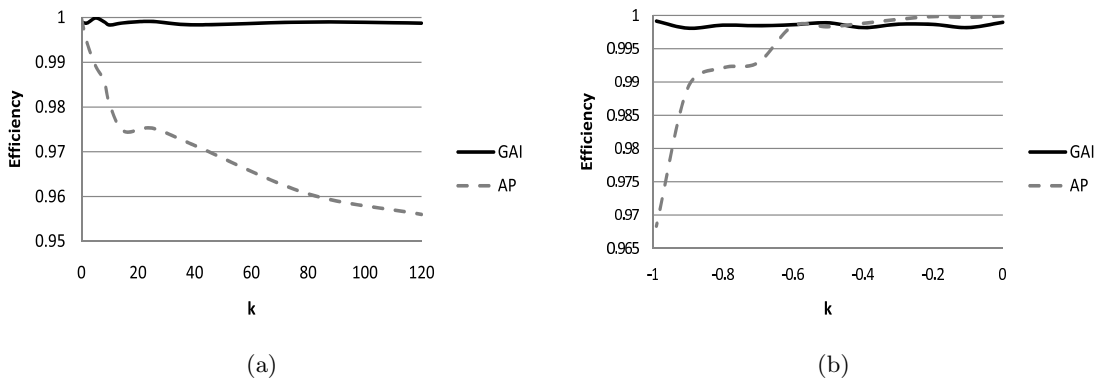

(a)                                                    (b)

Figure 3: (a) Efficiency as a function of $k \geq 0$ (complements). (b) Efficiency as a function of $k \leq 0$ (substitutes).

Figures 3a and 3b present efficiency as a function of the MUI-factor $k$, for complements and substitutes, respectively. We used a fixed GAI structure with four elements, the largest of which has four attributes, and imposed the same $k$ on all elements. As expected, the stronger the complementarity among the attributes, the lower the efficiency of AP, whereas this relationship does not affect the efficiency of GAI auctions. For the case of substitutes, in contrast, the additive approximation performs well, as efficiency starts to deteriorate only for extreme values of $k$. Very roughly, we can say that when relationship among attributes (within each GAI element) is limited to (mild) substitutions, it could be a good idea to use an additive approximation. Unfortunately, our interpretation of the parameter $k$ lacks quantitative scaling: there is no clear intuition of what the actual numbers mean, beyond the qualitative classification mentioned above.

To summarize this part, the experimental results show that GAI auctions yield significant efficiency improvement in comparison to an additive auction, on almost all classes of evaluations. Though the efficiency of an additive auction may come across as relatively high and perhaps sufficient, such an observation is misleading in several respects. (i) In large procurement events, 5–10% efficiency differences translate to large amounts of money. (ii) The wider efficiency loss of an additive auction (with no theoretical bound) may have an impact on incentives; SB may no longer be an approximate ex-post Nash equilibrium. (iii) Efficiency is expected to deteriorate for larger problems with larger GAI elements, and in particular if FOPI does not hold for many of the attributes. (iv) As argued in Section 7.3, we expect practical additive auctions to perform worse than AP with our tailored approximation.

## 8.2 Computational Analysis

The computational tasks required by auction simulations were performed using the algorithms described in Appendix C. These algorithms have been suggested and applied for combinatorial optimization problems before (Dechter, 1997; Nilsson, 1998), therefore the computational runtime to process a round is not of particular interest to this work. Instead, we focus on the number of rounds the auction requires. We tested the number of rounds required by auctions GAI and AP, under fully random and FOPI preferences, varying three of the parameters: $\xi$ (size of largest GAI element), $e$ (connectivity), and $\delta$.

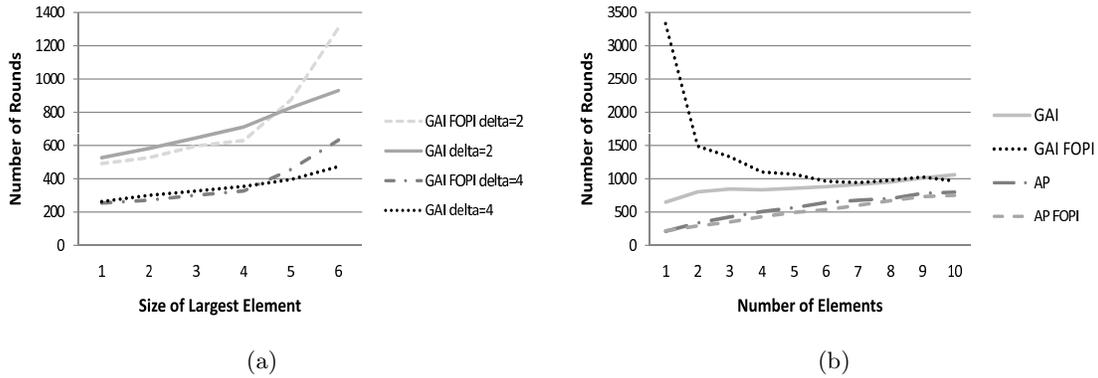

(a)                                        (b)

Figure 4: Number of rounds as a function of: (a) the size of largest GAI element ($\xi$), given $e = 5$, (b) the number of GAI elements ($e + 1$), given $\xi = 5$ and $\delta = 2$.

The complexity in terms of number of rounds is shown in Figure 4a (with respect to $\xi$) and Figure 4b (with respect to the number of elements). We observe that under FOPI the GAI auction takes much longer to converge, compared to the case of random preferences. The reason is that under FOPI, the sellers' and the buyer's preferences can in general be seen as opposites: at the same price, and for a specific attribute, the buyer prefers higher quality, whereas the sellers prefer lower quality (given fixed values for the rest of the attributes), and everyone agrees on the relative quality of attribute values. The apparent difference in the growth rate (the FOPI case seems to have a steeper curve) is somewhat misleading: for $\xi = 8$ (not shown) GAI under random preferences is already caught up with the same curve we see for the FOPI case. The number of rounds, as expected, grows exponentially with the size of the largest element. However, as observed from Figure 4b, this number does not grow quickly as a function of the number of elements, supporting the theoretical arguments of Section 6.2. Note also that the variance chosen for traders' preferences is fixed, thus for a small number of elements the variance over them is wider, resulting in the large number of rounds required by GAI FOPI in that case.

For AP, the only implication of increasing $\xi$ is the respective increase in the number of attributes. As a result, the complexity of AP (not shown) grows very slowly with the increase in $\xi$. For the FOPI case, with $\delta = 2$, AP takes an average of 481 rounds for $\xi = 1$ (6 attributes) and 546 rounds for $\xi = 6$ (19 attributes). The numbers are slightly higher for random preferences (523 to 628).

For high-dimensional multiattribute auctions, we expect that participation would typically be automated through software bidding agents (Wellman, Greenwald, & Stone, 2007). Under these circumstances, an auction taking up to thousands of rounds should not cause a concern. However, if for some reason rounds are expensive, we might reconsider adopt-ing additive auctions, and sacrifice efficiency in order to decrease the number of rounds. Alternatively, we could keep using GAI auctions and increase $\epsilon$ (and with it $\delta$). The final experiment compares these two alternatives. We vary the level of $\epsilon$, in order to view efficiency as a function of the number of rounds (Figure 5). The GAI structure used for this experiment has $e = 5$ and $\xi = 5$.

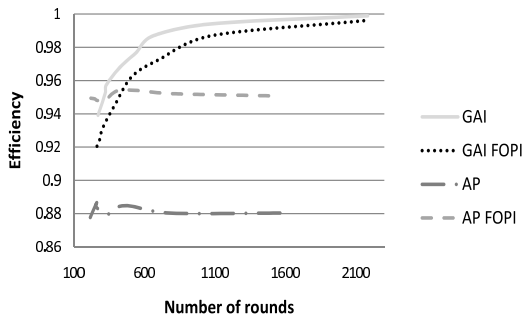

Figure 5: Efficiency as a function of the number of rounds.

As evident from the chart, in most cases GAI achieves better efficiency even for a fixed number of rounds. The only exception is when the budget of rounds is very small (under 200), and FOPI holds. In such case we need to pay with more rounds in order to get the higher efficiency.

The total computation time, carried out by a GAI auction with 10 elements, $\xi = 5$, $d = 3$, $\delta = 2$, and the rest of the parameters fixed as above, is around 11 seconds on average, using an Intel Dual Core (2.00 Ghz) CPU, with 2048 MB RAM.

### 8.3 Information Revelation

A key difference between the mechanism proposed here and most previous literature is in the extent to which the buyer is required to reveal preference information. In GAI auctions, the buyer does not need to reveal all of its private preference information up front. Of course, the price changes do reveal some of the buyer's information. Another experimental question is therefore whether this mechanism significantly reduces the overall amount of information revealed by the buyer.

PK study information revelation by both the buyer and the seller, under an additivity assumption. When the utility function is additive the amount of information revealed can be measured in terms of constraints on the linear weights. Sellers can infer bounds on the buyer's set of weights, and the amount of information hidden from them is represented by the fraction of the simplex that satisfies those constraints. This simplex analysis is not possible for GAI utilities. We suggest an alternative geared towards the kind of information revealed by the GAI auctions.

In GAI auctions, the buyer's private information is partially revealed through the selection of the buyer's preferred set $\mathcal{M}^t$. The auction does not need to announce this directly; in general the sellers can infer that a sub-configuration is in $\mathcal{M}^t$ only if it received a bid (usually sellers will observe this only for their own bids), yet its price does not change in the next round. We therefore measure exactly that—for how many sub-configurations $\theta^r$ there was at least one round $t$ such that $\theta^r \in \mathcal{M}^t \cap \mathcal{B}_i^t$ for some $i$. More specifically, we define such a sub-configuration as *revealed*, and within each GAI element we measure the fraction of sub-configurations that are revealed by the end of the auction. This measurement overestimates the information that is actually revealed, as sellers can infer some bounds on relative preferences but not the precise values of the functions $f_b(\cdot)$. Moreover, it assumes that each seller observes all bids (meaning that sellers share bid information with each other), an unrealistic event in practice.

Based on this criterion, GAI auctions reveal on average 15%–25% of the buyer's preferences when preferences exhibit FOPI, and 10%–15% when the subutilities are completely random. It does not seem to systematically depend on any other parameter we tested. This validates our claim as to the advantage that GAI auctions promise over second-score types of auctions.

## 9. Conclusions

We propose a novel exploitation of preference structure in multiattribute auctions. Rather than assuming full additivity, or no structure at all, we model preferences using the generalized additive independence (GAI) decomposition. We show how a GAI representation can be constructed from relatively simple statements of willingness-to-pay, and develop an iterative auction mechanism directly relying on the decomposition. Our auction mechanism generalizes the preference modeling employed by Parkes and Kalagnanam (2005), while in essence retaining their information revelation properties. It allows for a range of tradeoffs between the accuracy of preference representation and computational complexity of the auction, as well as the tradeoff between buyer information revelation and the number of rounds required for convergence.

We performed a simulation study of our proposed multiattribute auctions, compared to a mechanism that assumes additive preferences. The study validated the usefulness of GAI auctions when preferences are non-additive but GAI, and allowed us to quantify the advantages for specific classes of preferences. In general, we found that our design yields significantly higher economic efficiency in comparison to additive auctions. When the GAI subutilities exhibit internal structures, such as FOPI, the efficiency loss of additive approximation is less severe, but in most cases the benefit of an accurate GAI model is still significant. Using an additive approximation may be a reasonable approach when the GAI structure is fairly similar to an additive one, or when the auction must terminate within a small number of rounds.

The tradeoff between expressive and compactness of preference representation is ubiquitous in applications involving preferences. On one hand, we would like to ask users for as little as possible information; on the other, users' preference statements may not be accurate or even meaningful if they cannot express important dependencies. In such problems it could be useful to experimentally compare the accuracy of GAI and additive representations. The experimental methodologies used in this study, in particular the generation of random structured utility functions, and finding an additive approximation to GAI functions, may therefore prove applicable to a broader class of preference research problems in which this tradeoff exists.

## Acknowledgments

This work was supported in part by NSF grants IIS-0205435 and IIS-0414710, and the STIET program under NSF IGERT grant 0114368. Yagil Engel was supported in part by the Aly Kaufman fellowship at the Technion. We thank anonymous reviewers for many useful comments and suggestions.

## Appendix A. Proofs of Section 3.1

**Lemma 2.** *Let $u(A)$ be an MVF representing preference differences, and let $X, Y, Z$ define a partition of $A$. Then $\mathrm{CDI}(X, Y \mid Z)$ iff*

$$u(A) = u(X^0, Y, Z) + u(X, Y^0, Z) - u(X^0, Y^0, Z),$$

*for arbitrary instantiations $X^0, Y^0$.*

*Proof.* Let $X^0, Y^0$ be arbitrary instantiations.

$$u(X, Y, Z) = u(X, Y, Z) - u(X^0, Y, Z) + u(X^0, Y, Z) = u(X, Y^0, Z) - u(X^0, Y^0, Z) + u(X^0, Y, Z)$$

The second equality holds iff for any $X^0, Y^0$, $\mathrm{CDI}(X, Y \mid Z)$. □

**Theorem 3 (CDI-GAI Theorem).** *Let $G = (A, E)$ be a CDI map for $A$, and $\{I_1, \ldots, I_g\}$ a set of overlapping maximal cliques. Then*

$$u(A) = \sum_{r=1}^{g} f_r(I_r), \tag{A.1}$$

*where*

$$\begin{aligned}
f_1 &= u([I_1]), \text{ and} \tag{A.2}\\
\text{for } r = 2, \ldots, g, \quad f_r &= u([I_r]) + \sum_{j=1}^{r-1} (-1)^j \sum_{1 \le i_1 < \cdots < i_j < r} u([\bigcap_{s=1}^{j} I_{i_s} \cap I_r]).
\end{aligned}$$

*Proof.* We actually prove a somewhat stronger result.

**Claim.** *Let $G$ be a CDI map for utility function $u(\cdot)$. Let $\mathcal{Q} = \{C_1, \ldots, C_w\}$ denote the set of maximal cliques of $G$. Then,*

$$u(A) = \sum_{k=1}^{w} (-1)^{k+1} \sum_{1 \le i_1 < \cdots < i_k \le w} u([\bigcap_{s=1}^{k} C_{i_s}]). \tag{A.3}$$

Let $G^0 = (A, E^0)$ be the complete graph over the nodes of $G$. By definition of CDI map, each edge $(x, y) \in E^0 \setminus E$ implies $\mathrm{CDI}(x, y)$. We use induction on a series of edge removals. starting from the graph $G^0$, at each step $i$ we remove an edge in $E^0 \setminus E$ to get graph $G^i$. After the last step $i = |E^0| - |E|$ and $G^{|E^0| - |E|} = G$. We prove that the claim holds for each graph $G^i$. Since $A$ is the only clique in $G^0$, in step 0, $\mathcal{Q}^0 = \{A\}$ and the claim trivially hold. Following the process for step 1 provides intuition as for how the final decomposition is obtained. We pick a pair of nodes $(x, y)$ such that CDI$(x, y)$. We use the notation $S^{-a} = S \setminus \{a\}$ for any $S \subseteq A$ and $a \in A$. By Lemma 2 ,

$$
\begin{aligned}
u(A) &= u(x, y, A^{-x,y}) \\
&= u(x^0, y, A^{-x,y}) + u(x, y^0, A^{-x,y}) - u(x^0, y^0, A^{-x,y}) \\
&= u([A^{-x}]) + u([A^{-y}]) - u([A^{-x} \cap A^{-y}]).
\end{aligned}
\tag{A.4}
$$

The set of maximal cliques of $G^1$ is $\mathcal{Q}^1 = \{A^{-x}, A^{-y}\}$. Equation (A.4) shows that (A.3) holds for $\mathcal{Q}^1$.

For proving the induction step, we assume (A.3) holds at step $i$, and show they carry over to step $i+1$. Let $(x, y)$ denote the edge removed in step $i+1$. Let $\hat{C}_1, \ldots, \hat{C}_d$ (WLOG) indicate all the sets in $\mathcal{Q}^i$ that include both $x$ and $y$. Similar to (A.4), we observe that

$$
u([\hat{C}_1]) = u([\hat{C}_1^{-x}]) + u([\hat{C}_1^{-y}]) - u([\hat{C}_1^{-x} \cap \hat{C}_1^{-y}]).
\tag{A.5}
$$

Similarly for any $k = 1, \ldots, w^i - 1$, and $1 < i_1 < \cdots < i_k \leq w^i$,

$$
u([\bigcap_{s=1}^{k} \hat{C}_{i_s} \cap \hat{C}_1]) = u([\bigcap_{s=1}^{k} \hat{C}_{i_s} \cap \hat{C}_1^{-x}]) + u([\bigcap_{s=1}^{k} \hat{C}_{i_s} \cap \hat{C}_1^{-y}]) - u([\bigcap_{s=1}^{k} \hat{C}_{i_s} \cap \hat{C}_1^{-x} \cap \hat{C}_1^{-y}]).
\tag{A.6}
$$

In (A.3) (assumed to hold before this step) each term that includes $\hat{C}_1$ can be substituted according to (A.5) or (A.6). Doing so will result in (A.3) holding for the set $(\mathcal{Q}_i \setminus \{\hat{C}_1\}) \cup \{\hat{C}_1^{-x}, \hat{C}_1^{-y}\}$.

We repeat the same operation for $C_2, \ldots, C_d$, and define the resulting collection

$$
\mathcal{Q}^{i+1} = (\mathcal{Q}^i \setminus \{\hat{C}_1, \ldots, \hat{C}_d\}) \cup \{\hat{C}_1^{-x}, \hat{C}_1^{-y}, \ldots, \hat{C}_d^{-x}, \hat{C}_d^{-y}\}.
$$

All elements in $\mathcal{Q}^{i+1}$ are subsets of elements in $\mathcal{Q}^i$, which are all maximal cliques of $G^i$. We now verify the induction property:

- Any element in $\mathcal{Q}^{i+1}$ is a clique in $G^{i+1}$, because the only difference between $G^i$ and $G^{i+1}$ is the removed edge $(x, y)$, and no set in $\mathcal{Q}^{i+1}$ includes both $x$ and $y$.

- Any such clique in $C \in \mathcal{Q}^{i+1}$ is maximal, because it is a subset of a maximal clique of $\hat{C} \in G^i$, and either: (i) $y \in \hat{C}$ and $C = \hat{C} \setminus \{x\}$ or (ii) $x \in \hat{C}$ and $C = \hat{C} \setminus \{y\}$, or (iii)$C = \hat{C}$. $x$ and $y$ are no longer connected so $C$ remains maximal in all cases.

- If $M$ is a maximal clique in $G^{i+1}$, then $M \subseteq \hat{C}$ for some $\hat{C} \in \mathcal{Q}^i$. Again either $M = \hat{C}$, or $M = \hat{C} \setminus \{x\}$, or $M = \hat{C} \setminus \{y\}$, and in all three cases $M$ is an element in $\mathcal{Q}^{i+1}$.

This proves the induction step.

As a result, in the last step the decomposition (A.3) holds for the set $\mathcal{Q} = \mathcal{Q}^{|E_0| - |E|}$, which is the set of maximal cliques of $G$, and hence the claim is proved.

Now define $f_1, \ldots, f_g$ according to (A.2). By the claim, we get that (A.1) holds. $\qquad\square$

## Appendix B. Proofs of Section 6.1

### B.1 Proving Lemma 5

**Lemma 5.** *Let $\Psi$ be a set of configurations, all are within $\epsilon$ of maximizing profit for a trader $\tau$ at the a given price vector. Let $\Phi = \{\theta_r \mid \theta \in \Psi, r \in \{1, \ldots, g\}\}$. Then any consistent cover in $\Phi$ is within $g\epsilon$ of maximizing profit for $\tau$ at these prices.*

We show that given a suboptimal consistent cover $\theta$ over $\Phi$, we can find a suboptimal member in $\Psi$, contradicting the premise of the lemma. We do that by traversing the GAI tree in a depth-first manner, at each step we flip the sub-configurations corresponding to the elements of the subtree to a set of sub-configurations that have the same source configuration in $\Psi$ as the parent of that subtree (thus "trimming" that subtree). This, as we show, results in another consistent cover that is also sub-optimal. Eventually we obtain a configuration in $\Psi$ which is still suboptimal.

For that purpose we introduce the following notions:

- The operator $\oplus$ turns a set of sub-configurations, which is a consistent cover, into a configuration:
$$\oplus\{\theta_1, \ldots, \theta_g\} = (\theta_1, \ldots, \theta_g).$$

- Let $\theta$ be a consistent cover over $\Phi$. The **$\Psi$-source** of an element $\theta_r$ is a configuration $\hat{\theta} \in \Psi$ from which it originated (meaning, $\hat{\theta}_r = \theta_r$).

- The operation **trim** replaces some of the sub-configurations of a given configuration $\theta$ with a corresponding set of sub-configurations of a different configuration $\hat{\theta}$, according to the following rules. Let $\mu^i$ denote the indices of the GAI elements, corresponding to a subtree in the GAI-tree, whose root is the GAI element $I_i$. Let $\theta$ denote a consistent cover over $\Psi$. The operation $\Psi$-trim over $\theta$ and $\mu^i$ is defined if all the elements in $\theta$ corresponding to $\mu^i$ have the same $\Psi$-source. Formally, there exists $\hat{\theta} \in \Psi$, such that $\forall \theta_r$, if $r \in \mu^i$ then $\theta_i = \hat{\theta}_i$. Now, Let $\gamma$ be the parent of $I_i$, or an arbitrary element outside $\mu^i$ if $\mu^i$ is disconnected from the rest of the graph. Let $\hat{\theta} \in \Psi$ be the source of $\theta_\gamma$. Then
$$\Psi\text{-trim}(\mu^i, \theta) = \oplus\{\theta_r \mid r \notin \mu^i\} \cup \{\hat{\theta}_r \mid r \in \mu^i\}$$

That is we replace each of $\theta_r$ for $r \in \mu^i$ by the corresponding sub-configuration in $\hat{\theta}$, so that in the resulting configuration all the elements corresponding to $\mu^i$ have the same $\Psi$-source as the parent of $I_i$.

**Lemma B.1.** *$\theta' = \Psi\text{-trim}(\mu, \theta)$ is a consistent cover.*

*Proof.* We need to show that any pair of sub-configurations in the set $\{\theta_r \mid r \notin \mu^i\} \cup \{\hat{\theta}_r \mid r \in \mu^i\}$ are consistent, that is they assign the same value to any attribute that appear in both corresponding GAI elements.

The sub-configurations $\{\hat{\theta}_r \mid r \in \mu^i\}$ are internally consistent because they have a mutual $\Psi$-source $\hat{\theta}$. The sub-configurations $\{\theta_r \mid r \notin \mu^i\}$ are internally consistent because they are all sub-configurations of $\theta$. Let $r_1 \in \mu^i$ and $r_2 \notin \mu^i$ denote indices of GAI elements, such that $I_{r_1} \cap I_{r_2} \neq \emptyset$. Now, $I_{r_1}$ is in a subtree whose root is $I_i$, whereas $I_{r_2}$ is outside the subtree, so the path between them must go through $I_i$ and its parent $\gamma$. Due to the running intersection property of the GAI tree, $I_{r_1} \cap I_{r_2} \subseteq I_i \cap \gamma$. The corresponding sub-configurations $\hat{\theta}_i$ and $\theta_\gamma$ must be consistent because $\hat{\theta}$ is also the $\Psi$-source of $\gamma$, hence $\hat{\theta}_{r_1}$ and $\theta_{r_2}$ must also be consistent. $\qquad\square$

**Lemma B.2.** *Let $\Psi$ and $\Phi$ be defined as in Lemma 5, and let $\theta$ denote a consistent cover in $\Phi$. Then if $\theta' = \Psi\text{-}trim(\mu^i, \theta)$ (for some $i$), then $\pi_\tau(\theta') \leq \pi_\tau(\theta) + \epsilon$.*

*Proof.* Let $\tilde{\theta} \in \Psi$ denote the single $\Psi$-source of $\{\theta_r \mid r \in \mu^i\}$. Let $\mu = \mu^i$ and $\bar{\mu} = \{1, \ldots, g\} \setminus \mu$. If $\pi_\tau(\theta') > \pi_\tau(\theta) + \epsilon$, then (using Lemma 4)

$$\pi_\tau(\theta') = \pi_\tau(\theta'_\mu) + \pi_\tau(\theta'_{\bar{\mu}}) > \pi_\tau(\theta_\mu) + \pi_\tau(\theta_{\bar{\mu}}) + \epsilon,$$

and because $\theta_{\bar{\mu}} = \theta'_{\bar{\mu}}$,

$$\pi_\tau(\theta'_\mu) > \pi_\tau(\theta_\mu) + \epsilon.$$

Define the following cover:

$$\hat{\theta} = \oplus\{\theta'_r \mid r \in \mu\} \cup \{\tilde{\theta}_r \mid r \in \bar{\mu}\}$$

$\hat{\theta}$ is a consistent cover–again (as in Lemma B.1) the only possible intersection between an element from $\theta'$ and an element from $\tilde{\theta}$ is between $i$ (the root of $\mu = \mu^i$) and its parent $\gamma$. The corresponding sub-configurations $\theta'_i$ and $\tilde{\theta}_\gamma$ must be consistent for the following argument: $\tilde{\theta}_i$ is consistent with $\theta_\gamma$ because they appear together in $\theta$. $\theta'_i$ is consistent with $\theta_\gamma$ because they have the same $\Psi$-source by definition of $\Psi$-trim. Hence $\tilde{\theta}_i$ and $\theta'_i$ assign the same values to the attributes in $I_i \cap I_\gamma$. Now because $\tilde{\theta}_i$ is consistent with $\tilde{\theta}_\gamma$, so must be $\theta'_i$. We get

$$\pi_\tau(\hat{\theta}) = \pi_\tau(\theta'_\mu) + \pi_\tau(\tilde{\theta}_{\bar{\mu}}) > \pi_\tau(\theta_\mu) + \pi_\tau(\tilde{\theta}_{\bar{\mu}}) + \epsilon = \pi_\tau(\tilde{\theta}) + \epsilon.$$

The last equation follows from the fact that all sub-configurations of $\theta_\mu$ are from $\tilde{\theta}$. This contradicts $\epsilon$-optimality of $\tilde{\theta} \in \Psi$. $\qquad\blacksquare$

*Proof of Lemma 5.* Let $\theta^1$ be a consistent cover over $\Phi$ contradicting the lemma, meaning $\pi_\tau(\theta^1) \leq \max_{\theta \in \Theta} \pi_\tau(\theta) - g\epsilon$. We first reorder the GAI elements as $1, \ldots, g$, according to the order corresponding to backtracking in Depth-First-Search: that is, starting from the leftmost leaf, next move to its siblings, next their parent, and in general once all children of a node $I_i$ are visited, the next element to be visited is $I_i$. We perform a series of $g - 1$ $\Psi$-trim operations, resulting in a series $\theta^1, \ldots, \theta^g$. To do that, we must show that at each step $i$ the operation $\Psi$-trim$(\mu_i, \theta^i)$ is valid, that is the sub-configurations corresponding to $\mu_i$ have a mutual $\Psi$-source. If $I_i$ is a leaf, then $|\mu_i| = 1$ hence the elements of $\mu_i$ have a single source. Otherwise, $\theta^i$ is a result of trimming the subtrees of all children of $I_i$, hence by definition of $\Psi$-trim they all have the same $\Psi$-source as $\theta^i_i$.

Now, consider the resulting $\theta^g$. We assumed $\pi_\tau(\theta^1) < \max_{\theta \in \Theta} \pi_\tau(\theta) - g\epsilon$, hence by applications of Lemma B.2 in each of the $g - 1$ $\Psi$-trim operations, we get $\pi_\tau(\theta^g) < \max_{\theta \in \Theta} \pi_\tau(\theta) - \epsilon$. The last element $\theta^g$ is such that all its elements have a mutual $\Psi$-source, meaning $\theta^g \in \Psi$. Therefore, we got a contradiction to the $\epsilon$-optimality of $\Psi$. $\qquad\square$

## B.2 Proving Theorem 10

In order to prove Theorem 10 we need several additional claims.

**Lemma B.3.** *The price of at least one sub-configuration must be reduced in every round of phase A.*

*Proof.* In each round $t < T$ of phase A there exists an active seller $i$ for whom $B_i^t \cap M^t = \emptyset$. However to be active in round $t$, $B_i^t \neq \emptyset$. Let $\hat{\theta} \in B_i^t$. If $\forall r.\hat{\theta}_r \in \mathcal{M}^t$, then $\hat{\theta} \in M^t$ by definition of $M^t$. Therefore there must be $\hat{\theta}_r \notin \mathcal{M}^t$. $\qquad\square$

**Lemma B.4.** *The auction must reach phase B.*

*Proof.* By Lemma B.3 some prices must go down in every round of phase A. Lemma 7 ensures a lower bound on how much prices can be reduced during phase A, therefore the auction either terminates in phase A or must reach condition [SWITCH].

We set the initial prices high such that $\max_{\theta \in \Theta} \pi_b^1(\theta) < 0$, and then $\max_{\theta \in \Theta} \pi_b^t(\theta) < 0$ during phase A by Lemma 7. By Assumption $A_2$ the efficient allocation $(\theta^*, i^*)$ provides positive welfare, that is $\sigma_{i^*}(\theta^*) = \pi_b^t(\theta^*) + \pi_{i^*}^t(\theta^*) > 0$. $s_{i^*}$ is SB therefore she will leave the auction only when $\pi_{i^*}^t(\theta^*) < 0$. This can happen only when $\pi_b^t(\theta^*) > 0$, therefore $s_{i^*}$ does not drop in phase A. Because Phase A continues as long as at least one seller is active, the auction cannot terminate before reaching condition [SWITCH]. $\qquad\square$

Finally, the following lemma states that for any two sellers, the potential surplus of the first one to drop from the auction cannot be significantly higher than the potential surplus of the one that stayed longer.

**Lemma B.5.** *If sellers $s_i$ and $s_j$ are SB, and $s_i$ is active at least as long as $s_j$ is active in phase B, then*

$$\sigma_i(\eta_i) \geq \max_{\theta \in \Theta} \sigma_j(\theta) - (g+1)\epsilon.$$

*Proof.* From SB and the definition of phase B, $s_j$ drops when $\Delta > \pi_j^T(\eta_j)$. If $s_i$ did not drop before that point then $\pi_i^T(\eta_i) \geq \Delta - \epsilon > \pi_j^T(\eta_j) - \epsilon$. Because $\eta_i \in M^t$, we get from Corollary 6 that,

$$\pi_b^T(\eta_i) + \pi_i^T(\eta_i) \geq \max_{\theta \in \Theta} \pi_b^T(\theta) + \pi_j^T(\eta_j) - (g+1)\epsilon.$$

From Corollary 8, $\pi_j^T(\eta_j) = \max_{\theta \in \Theta} \pi_j^T(\theta)$. Therefore

$$\sigma_i(\eta_i) = \pi_b^T(\eta_i) + \pi_i^T(\eta_i) \geq \max_{\theta \in \Theta} \pi_b^T(\theta) + \max_{\theta \in \Theta} \pi_j^T(\theta) - (g+1)\epsilon \geq \max_{\theta \in \Theta} \sigma_j(\theta) - (g+1)\epsilon.$$

$\qquad\square$

**Theorem 10.** *Given a truthful buyer and SB sellers, the surplus of the final allocation is within $(g+1)\epsilon$ of the maximal surplus.*

*Proof.* From Lemma B.4 the auction terminates with an allocation $(s_i, \eta_i)$. From Lemma 9, the theorem is immediate in case the winning seller $s_i$ is the efficient seller. Otherwise the efficient seller is $s_j$ who dropped before or with $s_i$. The result is now immediate from Lemma B.5. $\qquad\square$

### B.3 Proving Theorem 12

We first adapt Lemma 7, Lemma 9, and Lemma B.5 to use $e + 1$ instead of $g$.

**Lemma B.6.** $\max_{\theta \in \Theta} \pi_b^t(\theta)$ *does not change in any round $t$ of phase A.*

*Proof.* Let $G$ be comprised of trees $G_1, \ldots, G_h$, let $\theta_j'$ denote the projection of a configuration $\theta'$ on the tree $G_j$, and let $g_j$ denote the number of GAI elements in $G_j$.

Assume there exists some $\theta_j'$ for which $\pi_b^{t+1}(\theta_j') > \pi_b^t(\theta_j')$. Then necessarily $p^{t+1}(\theta_j') = p^t(\theta_j') - \delta$. For that to happen it must be the case that some $w \leq g_j$ sub-configurations of $\theta_j'$ are not in $\mathcal{M}_j^t$, and $\delta = \frac{w\epsilon}{g}$. In that case, by definition of $\mathcal{M}_j^t$,

$$\pi_b^t(\theta_j') < \max_{\theta_j \in \Theta_j} \pi_b^t(\theta_j) - g_j \frac{\epsilon}{g}.$$

Therefore,

$$\pi_b^{t+1}(\theta_j') = \pi^t(\theta_j') + \delta = \pi^t(\theta_j') + \frac{w\epsilon}{g} \leq \pi^t(\theta_j') + \frac{g_j\epsilon}{g} < \max_{\theta_j \in \Theta_j} \pi_b^t(\theta_j).$$

This is true for any $\theta_j'$ whose profit improves, therefore $\max_{\theta_j \in \Theta_j} \pi_b^t(\theta_j)$ does not change during phase A. Now

$$\max_{\theta \in \Theta} \pi_b^t(\theta) = \max_{\theta \in \Theta} \sum_{j=1}^h \pi_b^t(\theta_j) = \sum_{j=1}^h \max_{\theta_j \in \Theta_j} \pi_b^t(\theta_j).$$

The last equality holds because the optimal values for disconnected components of the GAI tree are independent of each other. As a result, $\max_{\theta \in \Theta} \pi_b^t(\theta)$ as well does not change during phase A. $\square$

**Lemma B.7.** *For SB seller $s_i$, $\eta_i$ is $(e + 1)\epsilon$-efficient:*

$$\sigma_i(\eta_i) \geq \max_{\theta \in \Theta} \sigma_i(\theta) - (e + 1)\epsilon.$$

The proof is identical to the proof of Lemma 9, replacing $g$ by $e+1$ and using Corollary 11 instead of Corollary 6.

**Lemma B.8.** *If sellers $s_i$ and $s_j$ are SB, and $s_i$ is active at least as long as $s_j$ is active in phase B, then*

$$\sigma_i(\eta_i) \geq \max_{\theta \in \Theta} \sigma_j(\theta) - (e + 2)\epsilon.$$

The proof here too is identical to the proof of Lemma B.5, using Corollary 11 instead of Corollary 6.

**Theorem 12.** *Given a truthful buyer and SB sellers, the surplus of the final allocation is within $(e + 2)\epsilon$ of the maximal surplus.*

*Proof.* The proof is identical to the proof of Theorem 10, replacing Lemmas 9 and B.5 with lemmas B.7 and B.8, respectively. $\square$

### B.4 Lemma 13 and Theorem 14

**Lemma 13.** *When sellers are SB, the GAI auction payment is sell-side* $(e + 2)\epsilon$*-VCG.*

*Proof.* Trivially, we consider only the winning seller $s_i$. In the case that the final price is above buyer's valuation the payment $u_b(\eta_i)$ is exactly the VCG payment. We can therefore assume that the final price is not above the buyer's valuation, and the payment to the winning seller is $p^T(\eta_i) - \Delta$. Let $s_j$ be the second best seller. $s_j$ drops before $s_i$, when the discount is $\Delta - \epsilon$, hence,

$$\Delta + \epsilon > \pi_j^T(\eta_j) = \max_{\theta \in \Theta} \pi_j^T(\theta). \tag{B.1}$$

From Corollary 6,

$$u_b(\eta_i) - p^T(\eta_i) \geq \max_{\theta \in \Theta} \pi_b^T(\theta) - (e + 1)\epsilon.$$

Therefore (using (B.1) for the second inequality)

$$p^T(\eta_i) - \Delta \leq u_b(\eta_i) - \max_{\theta \in \Theta} \pi_b(\theta) + (e + 1)\epsilon - \Delta <$$
$$u_b(\eta_i) - \max_{\theta \in \Theta} \pi_b^T(\theta) + (e + 2)\epsilon - \max_{\theta \in \Theta} \pi_j^T(\theta) \leq u_b(\eta_i) - \max_{\theta \in \Theta} \sigma_j(\theta) + (e + 2)\epsilon. \tag{B.2}$$

Bow because $s_j$'s survived in the auction until the discount was $\Delta - \epsilon$,

$$\Delta \leq \pi_j^T(\eta_j) + \epsilon.$$

Meaning:

$$p^T(\eta_j) - \Delta \geq c_j(\eta_j) - \epsilon. \tag{B.3}$$

From Corollary 6

$$u_b(\eta_i) - p^T(\eta_i) \leq u_b(\eta_i) - p^T(\eta_j) + (e + 1)\epsilon.$$

Therefore (using (B.3) for the second inequality)

$$p^T(\eta_i) - \Delta \geq u_b(\eta_i) - u_b(\eta_j) + p^T(\eta_j) - (e + 1)\epsilon - \Delta \geq$$
$$u_b(\eta_i) - (u_b(\eta_j) - c_j(\eta_j)) - \epsilon - (e + 1)\epsilon \geq u_b(\eta_i) - \max_{\theta \in \Theta} \sigma_j(\theta) - (e + 2)\epsilon. \tag{B.4}$$

Equations (B.2) and (B.4) place the payment $p^T(\eta_i) - \Delta$ within $(e + 2)\epsilon$ from $s_i$'s VCG payment. □

**Theorem 14.** *SB is a* $(3e + 5)\epsilon$ *ex-post Nash equilibrium for sellers in the GAI auction. That is, sellers cannot gain more than* $(3e + 5)\epsilon$ *by deviating from SB, given that other sellers follow SB.*

Let $s_1$ play some arbitrary strategy $\rho_1$ against SB sellers $s_2, \ldots, s_n$. If $s_1$ does not win she would clearly have done no worse using SB, therefore we assume $s_1$ wins $\eta_1$ in final price $\tilde{p}$ and that she gains at least $(3e + 5)\epsilon$ from the trade. Let $i \in 2, \ldots n$. The calculation of (B.2) assumed nothing on the winning trader's strategy, therefore it applies here as well:

$$\tilde{p} = p^T(\eta_1) - \Delta \leq u_b(\eta_1) - \max_{\theta} \sigma_i(\theta) + (e + 2)\epsilon. \tag{B.5}$$

Next, define the following cost function: $\hat{c}_1(\eta_1) = \tilde{p} - (2e + 3)\epsilon$ and $\hat{c}_1(\theta') = \infty, \forall \theta' \neq \eta_1$. Assume $s_1$ plays SB for $\hat{c}_1$.

**Claim.** *By playing SB assuming cost $\hat{c}_1$, $s_1$ is still the winner, and her profit (wrt to $c_1(\cdot)$) is within $(2e+3)\epsilon$ of her profit playing $\rho_1$.*

*Proof.* Clearly, $s_1$ bids only on $\eta_1$. Let $\hat{p}(\cdot)$ denote prices in the end of phase A in the new instance of the auction, let $\hat{\pi}_b(\cdot)$ denote the buyer's profit, and let $\hat{\Delta}$ be the final discount. Now assume for a moment that prices reach $s_1$'s limit, that is $\hat{\Delta} = \pi_1(\eta_1) = \hat{p}(\eta_1) - \hat{c}_1(\eta_1) = \hat{p}(\eta_1) - (\tilde{p} - (2e+3)\epsilon)$.

Now (for the inequality, use $\hat{p}^T(\eta_1) = u_b(\eta_1) - \hat{\pi}_b^T(\eta_1)$ and also (B.5)),

$$\hat{\Delta} = \hat{p}^T(\eta_1) - \tilde{p} + (2e+3)\epsilon > u_b(\eta_1) - \hat{\pi}_b^T(\eta_1) - (u_b(\eta_1) - \max_{\theta} \sigma_i(\theta) + (e+2)\epsilon) + (2e+3)\epsilon$$

$$= \max_{\theta} \sigma_i(\theta) - \hat{\pi}_b^T(\eta_1) + (e+1)\epsilon.$$

(B.6)

Let $\hat{\eta}_i$ denote the configuration chosen for some seller $s_i$ at the end of phase A in the new instance. Since $\hat{\eta}_i \in M^T$ in that instance, we get that $\hat{\pi}_b^T(\hat{\eta}_i) \geq \hat{\pi}_b^T(\eta_1) - (e+1)\epsilon$. Therefore we can modify (B.6) to state,

$$\hat{\Delta} > \sigma_i(\hat{\eta}_i) - \hat{\pi}_b^T(\hat{\eta}_i) = \hat{p}^T(\hat{\eta}_i) - c_i(\hat{\eta}_i),$$

(B.7)

meaning that if prices reached the limit of $s_1$, all the other sellers dropped off. That shows that $s_1$ wins in the new instance as well. Furthermore, the lowest possible price paid to $s_1$ is determined by $\hat{\Delta} = \hat{p}(\eta_1) - (\tilde{p} - (2e+3)\epsilon)$, hence that price is at least $\tilde{p} - (2e+3)\epsilon$. $\qquad \square$

*Proof of Theorem 14.* From Lemma 13:

$$\hat{p} \leq VCG(\hat{c}_1, c_2, \ldots, c_n) + (e+2)\epsilon.$$

Truthful reporting is a dominant strategy for sellers in one-sided VCG auctions. Therefore

$$VCG(\hat{c}_1, c_2, \ldots, c_n) \leq VCG(c_1, c_2, \ldots, c_n).$$

With the result of the claim we get

$$\tilde{p} \leq \hat{p} + (2e+3)\epsilon \leq VCG(c_1, c_2, \ldots, c_n) + (3e+5)\epsilon.$$

Therefore by playing $\rho_1$, $s_1$ could not have gained more than $(3e+5)\epsilon$ above her worst-case payoff for playing SB with respect to her true cost $c_1$. $\qquad \blacksquare$

## Appendix C. Proofs of Section 6.2

**Theorem 15.** *The computation of $\mathcal{M}^t$ can be performed in time $O(g|\mathcal{I}|^2)$. Moreover, the total time spent on this task throughout the auction is $O(g|\mathcal{I}|(|\mathcal{I}| + T))$.*

*Proof.* For simplicity of notations we assume that there is a single (connected) GAI-tree. The extension to multiple connected components is immediate because each $M_j^t$ is computed separately.

The functions $u_b$ and $p^t$ have the same GAI form, hence the function $\pi_b^t = u_b - p^t$ has the same GAI form. As have been noted before (Boutilier et al., 2001), functions in GAI form can be optimized using variable elimination schemes in cost networks (Dechter, 1997). In fact, our GAI structure is already a tree, in which case the optimization is linear in the size of the domain which is $|\mathcal{I}|$. However, $\mathcal{M}^t$ includes sub-configurations of *all* configurations within $\epsilon$ of $\max_\theta \pi_b(\theta)$. To find it, we must find the maximum of $\pi_b^t$, add its sub-configurations to $\mathcal{M}^t$, then find the best configuration which is not already in $M^t$ (that is, maximal in $\Theta \setminus M^t$) and so on. This can be done by the following procedure, adapted from the work of Nilsson (1998):

1. For i=1,..., g:

   - Define $\Theta_i = \{\theta \in \Theta \mid \theta_1, \ldots, \theta_{i-1} \in \mathcal{M}^t \text{ and } \theta_i \notin \mathcal{M}^t\}$.
   - Find $\theta^i = \arg\max_{\theta \in \Theta_i} \pi_b^t(\theta)$.

2. The best configuration in $\Theta \setminus M^t$ is $\theta^* = \arg\max_{i=1,\ldots,g} \pi_b^t(\theta^i)$ (which means, a configuration which has at least one sub-configuration not in $\mathcal{M}^t$).

If $\pi_b^t(\theta^*) \geq \max_{\theta \in \Theta} \pi_b^t(\theta) - \epsilon$, then each sub-configuration of $\theta^*$ that is not already in $\mathcal{M}^t$ is added to $\mathcal{M}^t$. Otherwise, $\mathcal{M}^t$ is ready.

The procedure itself performs $g$ optimizations, each takes linear in the size of the domain. This amounts to $O(g|I|)$. Each time this procedure is done, either at least one sub-configuration is added to $\mathcal{M}^t$, or $\mathcal{M}^t$ is ready. Therefore the number of times the procedure is done per round is bounded by the number of sub-configurations $|\mathcal{I}|$ plus one, giving the $O(g|\mathcal{I}|^2)$ bound. Moreover, $\mathcal{M}^t$ is monotonically increasing in the auction. In each round, we start from the $\mathcal{M}^t$ computed in the previous round. Throughout the auction, each application of the procedure either yields a new sub-configuration in $\mathcal{M}^t$, or terminates the round, so the total number of times the procedure is performed throughout the auction is bounded by $|\mathcal{I}| + T$, leading to the overall bound of $O(g|\mathcal{I}|(|\mathcal{I}| + T))$. $\qquad\square$

## Appendix D. Relating the MUI condition to Complements and Substitutes

The definitions for the *utility independence* (UI) condition and MUI can be found elsewhere (Keeney & Raiffa, 1976).

**Definition 19.** A MUI-factor of a set $A$ of MUI attributes is a solution to

$$1 + k = \prod_{i=1}^{n}(1 + kk_i).$$

Keeney and Raiffa (1976) (KR) show that there is at most one MUI-factor in addition to zero (Appendix 6B of their text). This ensures the soundness of the following adaptation to their MUI representation theorem:[9]

**Theorem D.1.** *Let $A$ be a set of MUI attributes.*

1. *If the only MUI-factor of $A$ is zero, then $u(A) = \sum_{i=1}^{n} k_i u_i(a_i)$.*

---

9. The theorem is adapted from the book of Keeney and Raiffa (1976), Theorem 6.1, page 289.

   *2. Otherwise, let $k \neq 0$ be a MUI-factor. Then*

$$u(A) = \frac{\prod_{i=1}^{n}[kk_i u_i(a_i) + 1] - 1}{k}. \tag{D.1}$$

   KR go on to point out that if $k > 0$ we can define $u'(A) = 1 + ku(A)$, a strategically equivalent function to $u(\cdot)$, and turn (D.1) into a multiplicative representation. This can be done in a similar fashion for $k < 0$. Further, they show that if MUI is known to exist, one elicitation query is sufficient in order to determine whether the form of the function is additive or multiplicative.

   The following relationship allow us to interpret the MUI factor with respect to complements and substitutes. The result generalizes and formalizes an intuition given by KR for the case of MUI between two attributes.

**Theorem D.2.** *Let $A$ be a set of MUI attributes, such that there is a MUI-factor $k \neq 0$. Then $k > 0$ iff all pairs of attributes in $A$ are complements, and $k < 0$ iff all pairs of attributes in $A$ are substitutes.*

*Proof.* The proof is based on the work of Keeney and Raiffa (1976), Theorem 6.1, as explained below.

   Assume that $u(\cdot)$ is normalized such that $u(A^0) = 0$. For each attribute $a \in A$, let $\overline{a} = \overline{\{a\}}$, and we know $UI(\overline{a}, a)$. Utility independence of such form leads to the following functional form: there exist functions $f$ and $g$ such that,

$$u(A) = f(a) + g(a)u(\overline{a}, a^0)$$

We instantiate this form with the assignment $\overline{a}^0$ and get

$$u(\overline{a}^0, a) = f(a) + g(a)u(\overline{a}^0, a^0) = f(a)$$

Hence $f(a) = u(\overline{a}^0, a)$, and $g(a) = \frac{u(A) - u(\overline{a}^0, a)}{u(\overline{a}, a^0)}$ (this development is done by KR). With $u(A^0) = 0$, we get

$$g(a) = \frac{u(A) - u(\overline{a}^0, a)}{u(\overline{a}, a^0) - u(\overline{a}^0, a^0)}. \tag{D.2}$$

In proof of Theorem 6.1, KR define the MUI-factor as follows:

$$k = \frac{g(a) - 1}{u(\overline{a}^0, a)}$$

The denominator is always positive. Furthermore, as shown by (D.2), when $g(a) > 1$, $u(A) - u(\overline{a}^0, a) > u(\overline{a}, a^0) - u(\overline{a}^0, a^0)$. In particular it means that for any $b \in \overline{a}$, $a$ and $b$ are complements, because the inequality holds when holding fixed all attributes in $\overline{a}$ but $b$. Similarly, when $g(a) < 1$, $a$ and any $b \in \overline{a}$ are substitutes. Putting these pieces together, we get the desired result. □

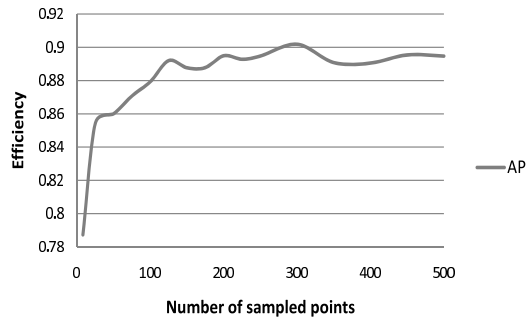

Figure 6: Efficiency of AP as a function of the number of sampling points used to devise the additive approximation.

## Appendix E. Optimal Regression Using a Small Sample

We show an experiment supporting the claim in Section 7.3: a larger set of sampling points than the one we used for the linear regression of the utility function cannot improve the efficiency of AP. Figure 6 shows the efficiency of AP as a function of the number of sampling points used, for the largest domain we used in the experiments: 25 attributes with $d = 4$ ($e = 9$ and $\xi = 5$). Similar results were shown for other distributions and for FOPI preferences. This chart is a result of 150 experiments for each of 10 points on the x-axis, the largest number of tests we used.